%% file: main.tex
\documentclass[11pt, a4paper, oneside]{Thesis} % Paper size, default font size and one-sided paper
\usepackage{wrapfig}
\usepackage{lscape}
\usepackage{rotating}
\usepackage{amsmath,amsfonts,bbm}
\usepackage{graphicx, caption}% for subfigure
\usepackage[]{algorithm2e}%for algorithm
\usepackage{float}% for H in figure
\usepackage[breaklinks=true]{hyperref}
\usepackage{breakcites}
\graphicspath{{Pictures/}} % Specifies the directory where pictures are stored
\usepackage[square, numbers]{natbib} % Use the natbib reference package - read up on this to edit the reference style; if you want text (e.g. Smith et al., 2012) for the in-text references (instead of numbers), remove 'numbers' v

\hypersetup{urlcolor=black, colorlinks=true} % Colors hyperlinks in blue - change to black if annoyingv`	
\title{\ttitle} % Defines the thesis title - don't touch this

\begin{document}
\makeatletter
\renewcommand*{\NAT@nmfmt}[1]{\textsc{#1}}
\makeatother

% prints author names as small caps

\frontmatter % Use roman page numbering style (i, ii, iii, iv...) for the pre-content pages

\setstretch{1.6} % Line spacing of 1.6 (double line spacing)

% Define the page headers using the FancyHdr package and set up for one-sided printing
\fancyhead{} % Clears all page headers and footers
\rhead{\thepage} % Sets the right side header to show the page number
\lhead{} % Clears the left side page header

\pagestyle{fancy} % Finally, use the "fancy" page style to implement the FancyHdr headers

\newcommand{\HRule}{\rule{\linewidth}{0.5mm}} % New command to make the lines in the title page

% PDF meta-data
\hypersetup{pdftitle={\ttitle}}
\hypersetup{pdfsubject=\subjectname}
\hypersetup{pdfauthor=\authornames}
\hypersetup{pdfkeywords=\keywordnames}

%----------------------------------------------------------------------------------------
%	TITLE PAGE
%----------------------------------------------------------------------------------------

\begin{titlepage}
\begin{center}

%\HRule \\[0.4cm] % Horizontal line
\vspace{0.4cm}
{\huge \bfseries \ttitle}\\[0.4cm] % Thesis title
\vspace{1.5cm}
%\HRule \\[1.5cm] % Horizontal line
 
\large \textit{A thesis submitted in fulfilment of the requirements\\ for the degree of \degreename}\\[0.3cm] % University requirement text
\textit{by}\\[0.4cm]

\href{https://github.com/kritiksoman}{\authornames}

\vfill
\graphicspath{ {./Figures/} }
\begin{figure}[hb]
  \centering
  \includegraphics[width=0.4\linewidth]{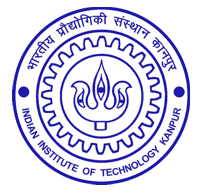}
\end{figure}

\DEPTNAME\\ % Research group name and department name
\textsc{ \UNIVNAME}\\[1.5cm] % University name
\large June 2020\\[2cm] % Date

\end{center}
\setcounter{page}{1}% Start page number with 1

\end{titlepage}

%----------------------------------------------------------------------------------------
%	DECLARATION PAGE
%	Your institution may give you a different text to place here
%----------------------------------------------------------------------------------------
\setcounter{page}{2}% Start page number with 1

\Declaration{{\vspace{1em}} % Add a gap in the Contents, for aesthetics
 
It is certified that the work contained in this thesis entitled ''\ttitle'' by \authornames has been carried out under my supervision and that it has not been submitted elsewhere for a degree.
\\[2cm]

\begin{minipage}{0.4\textwidth}
	\begin{flushleft} \large
		\emph{\large June 2020}\\[2cm] % Date
	\end{flushleft}
\end{minipage}
\begin{minipage}{0.65\textwidth}
	{\begin{center} \large
		\supname
		\begin{center}
			{Professor\\ 
			\normalsize{\deptname\\
			\univname}}
		\end{center}
	\end{center}}
\end{minipage}
\vfill{}}

\clearpage % Start a new page

%----------------------------------------------------------------------------------------
\thispagestyle{plain}

\begin{center}{\huge\bf Declaration\par}\end{center}
\normalsize
\vfil\vfil\null
%\cleardoublepage

{{\vspace{1em}} 
	
	This is to certify that the thesis titled "Semantic Mobile Base Station Placement" has been authored by me. It presents the research conducted by me under the supervision of Prof. K.S. Venkatesh. To the best of my knowledge, it is an original work, both in terms of research content and narrative, and has
	not been submitted elsewhere, in part or in full, for a degree. Further, due credit has been attributed to the
	relevant state-of-the-art and collaborations (if any) with appropriate citations and acknowledgements, in line
	with established norms and practices. 
	
	\begin{minipage}{0.4\textwidth}
		\begin{flushleft} \large
			\emph{\large June 2020}\\[2cm] % Date
		\end{flushleft}
	\end{minipage}
	\begin{minipage}{0.65\textwidth}
		{\begin{center} \large
				\vspace{4cm}
				Name: Kritik Soman\\
				Programme: M.Tech.\\
				\deptname\\
				\univname\\
				Kanpur 208016

		\end{center}}
	\end{minipage}
	\vfill{}}

\clearpage % Start a new page
%----------------------------------------------------------------------------------------
%	ABSTRACT PAGE
%----------------------------------------------------------------------------------------

%\addtotoc{Abstract} % Add the "Abstract" page entry to the Contents

\abstract{{\vspace{1em}} % Add a gap in the Contents, for aesthetics
Location of Base Stations in mobile networks plays an important role in coverage and signal strength received by users. As Internet of Things (IoT), autonomous vehicles and smart cities evolve, wireless network coverage will have a quintessential role in ensuring seamless connectivity. Due to use of higher carrier frequencies, blockages such as buildings and vegetation cause the communication to primarily be Line of Sight (LoS), further increasing the importance of base station placement. In this thesis, we propose a novel placement pipeline in which we perform semantic segmentation of aerial drone imagery using DeepLabv3+ and create its 2.5D model with the help of Digital Surface Model (DSM). This is used along with Vienna 5G simulator for finding the best location for deploying base stations by formulating the problem as a multi-objective function and solving it using Non-Dominated Sorting Genetic Algorithm II (NSGA-II). The case with and without prior deployed base station is considered. We then perform an evaluation of the base station deployment based on Signal to Interference Noise Ratio (SINR) coverage probability and user down-link throughput. This is followed by comparison with other base station placement methods and benefit offered by our approach.\\
\\

}

\clearpage % Start a new page

% %----------------------------------------------------------------------------------------
% %	ACKNOWLEDGEMENTS
% %----------------------------------------------------------------------------------------

% \setstretch{1.3} % Reset the line-spacing to 1.3 for body text (if it has changed)

% \acknowledgements{{\vspace{1em}} % Add a gap in the Contents, for aesthetics
% This thesis would not have been possible without the support and guidance of the people around me. I would like to express my sincere gratitude to my thesis supervisor Prof. K. S. Venkatesh for his constant support and motivation. His words of encouragement and guidance have helped me through many obstacles during my thesis.  I feel privileged to have worked under his guidance.

% I would also like to thank my friends, department staff members, my parents and brother Dr. Sumit Soman for supporting me throughout this journey.
% \\
% \\
% \flushright(Kritik Soman)

% }
% \clearpage % Start a new page

%----------------------------------------------------------------------------------------
%	LIST OF CONTENTS/FIGURES/TABLES PAGES
%----------------------------------------------------------------------------------------

\pagestyle{fancy} % The page style headers have been "empty" all this time, now use the "fancy" headers as defined before to bring them back

\lhead{\emph{Contents}} % Set the left side page header to "Contents"
\tableofcontents % Write out the Table of Contents

\lhead{\emph{List of Figures}} % Set the left side page header to "List of Figures"
\listoffigures % Write out the List of Figures

\lhead{\emph{List of Tables}} % Set the left side page header to "List of Tables"
\listoftables % Write out the List of Tables

%----------------------------------------------------------------------------------------
%	ABBREVIATIONS
%----------------------------------------------------------------------------------------

\clearpage % Start a new page

\setstretch{1.5} % Set the line spacing to 1.5, this makes the following tables easier to read

\lhead{\emph{Abbreviations}} % Set the left side page header to "Abbreviations"
\listofsymbols{ll} % Include a list of Abbreviations (a table of two columns)
{
\textbf{CDF}	& 	Cummulative Distribution Function \\
\textbf{CCDF}	& 	Complementary Cummulative Distribution Function \\
\textbf{CNN}	& 	Convolutional Neural Network \\
\textbf{CPU}	& 	Central Processing Unit \\
\textbf{DCNN}	& 	Deep Convolutional Neural Network \\
\textbf{DSM}	& 	Digital Surface Model \\
\textbf{FCN}	& 	Fully Convolutional Network \\
\textbf{FN}	& 	False Negative \\
\textbf{FP}	& 	False Positive \\
\textbf{GA}	& 	Genetic Algorithm \\

\textbf{GPU}	& 	Graphics Processing Unit \\
\textbf{HPBW}	& 	Half Power Beam Width \\
\textbf{IoT} 	& 	Internet of Things  \\
\textbf{IoU} 	& 	Intersection over Union  \\
\textbf{ISPRS}	& 	International Society for Photogrammetry and Remote Sensing \\
\textbf{LoS}	& 	Line of Sight \\
\textbf{LTE}	& 	Long Term Evolution \\
\textbf{MMW}	& 	Milli-Meter Waves \\
\textbf{MUS}	& 	Most Unserved Sector \\

\textbf{NSGA}	& 	Non-Dominated Sorting Genetic Algorithm \\
\textbf{PDF}	& 	Probability Distribution Function \\
\textbf{QoS}	& 	Quality of Service\\

\textbf{RF}	& 	Radio Frequency \\
\textbf{RGB}	& 	Red-Green-Blue \\
\textbf{RSRP}	& 	Reference Signal Received Power \\
\textbf{RTP}	& 	Reception Test Points \\
\textbf{SEAMO}	& 	Simple Evolutionary Algorithm for Multiobjective Optimization \\
\textbf{SINR}	& 	Signal to Interference Noise Ratio \\
\textbf{TN}	& 	True Negative \\
\textbf{TP}	& 	True Positive \\

\textbf{UAV}	& 	Unmanned Aerial Vehicle\\
\textbf{WSN}	&  Wireless Sensor Network \\
}

%----------------------------------------------------------------------------------------
%	PHYSICAL CONSTANTS/OTHER DEFINITIONS
%----------------------------------------------------------------------------------------
%
%\clearpage % Start a new page
%
%\lhead{\emph{Physical Constants}} % Set the left side page header to "Physical Constants"
%
%\listofconstants{lrcl} % Include a list of Physical Constants (a four column table)
%{
%Speed of Light & $c$ & $=$ & $2.997\ 924\ 58\times10^{8}\ \mbox{ms}^{-\mbox{s}}$ (exact)\\
%% Constant Name & Symbol & = & Constant Value (with units) \\
%}

%----------------------------------------------------------------------------------------
%	SYMBOLS
%----------------------------------------------------------------------------------------

%\clearpage % Start a new page
%
%\lhead{\emph{Symbols}} % Set the left side page header to "Symbols"
%
%\listofnomenclature{lll} % Include a list of Symbols (a two column table)
%{
%$p_c$ & Crossover probability \\
%$p_m$ & Mutation probability \\
%
%}

%----------------------------------------------------------------------------------------
%	DEDICATION
%----------------------------------------------------------------------------------------
%
\setstretch{1.3} % Return the line spacing back to 1.3
\pagestyle{empty} % Page style needs to be empty for this page
\dedicatory{Dedicated to my parents} % Dedication text
\addtocontents{toc}{\vspace{2em}} % Add a gap in the Contents, for aesthetics

%----------------------------------------------------------------------------------------
%	THESIS CONTENT - CHAPTERS
%----------------------------------------------------------------------------------------

\mainmatter % Begin numeric (1,2,3...) page numbering

\pagestyle{fancy} % Return the page headers back to the "fancy" style

% Include the chapters of the thesis as separate files from the Chapters folder
% Uncomment the lines as you write the chapters

\input{Chapters/Chapter1}
\input{Chapters/Chapter2} 
\input{Chapters/Chapter3}
\input{Chapters/Chapter4} 
\input{Chapters/Chapter5}

\input{Chapters/Chapter6} 
%\input{Chapters/Chapter7} 

%----------------------------------------------------------------------------------------
%	THESIS CONTENT - APPENDICES
%----------------------------------------------------------------------------------------

%\addtocontents{toc}{\vspace{2em}} % Add a gap in the Contents, for aesthetics
%
%\appendix % Cue to tell LaTeX that the following 'chapters' are Appendices
%
%% Include the appendices of the thesis as separate files from the Appendices folder
%% Uncomment the lines as you write the Appendices
%
%%\input{Appendices/AppendixA}
%%\input{Appendices/AppendixB}
%%\input{Appendices/AppendixC}
%
%\addtocontents{toc}{\vspace{2em}} % Add a gap in the Contents, for aesthetics

\backmatter

%----------------------------------------------------------------------------------------
%	BIBLIOGRAPHY
%----------------------------------------------------------------------------------------
\nocite{*}
\label{Bibliography}

\lhead{\emph{Bibliography}} % Change the page header to say "Bibliography"

\bibliographystyle{unsrt} % Use the "custom" BibTeX style for formatting the Bibliography

\bibliography{Bibliography} % The references (bibliography) information are stored in the file named "Bibliography.bib"

\end{document}

%% file: Chapters/Chapter1.tex
% Chapter Template

\chapter{Introduction} % Main chapter title

\label{Chapter1} % Change X to a consecutive number; for referencing this chapter elsewhere, use \ref{ChapterX}

\lhead{Chapter 1. \emph{Introduction}} % Change X to a consecutive number; this is for the header on each page - perhaps a shortened title

%----------------------------------------------------------------------------------------
%	SECTION 1
%----------------------------------------------------------------------------------------

\section{Scope and Objectives}
Mobile communication technologies have become the predominant mode of connectivity for present day devices such as mobiles, tablets, wireless sensor nodes, and Internet of Things (IoT) devices. This has triggered widespread deployment of mobile base stations all over the world. With every generation of wireless technology, the density of base stations per unit area has been increasing due to increase in carrier frequency. The effect of using a higher carrier frequency has also been illustrated in Fig. \ref{fig:f_compare} below. 

\begin{figure}[H]
	\centering
	\includegraphics[width=0.4\textwidth]{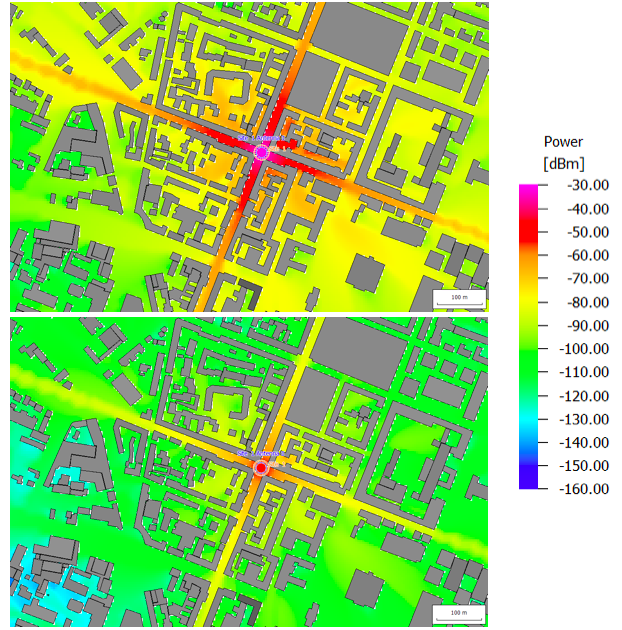}
	\caption{Effect of varying carrier frequency on received power for a single Base Station shown on the map of Munich. Top: Carrier frequency = 2.9 GHz, Bottom: Carrier frequency = 28 GHz}
	\label{fig:f_compare}
\end{figure}

As a result, the deployment strategy has transitioned from hexagonal cells to random deployment in voronoi like cells based on user density \cite{hao2017wireless}. In upcoming generations of wireless techonology like 5G with millimeter waves, the shape of cells have become so random that the difference between the theoretical and measured cell boundary is modelled with a Probability Distribution Function (PDF). The PDF has been experimentally shown to have self similarity properties \cite{ge2016wireless} like fractals and is referred to as wireless fractal phenomenon. The statistical properties of such irregular shaped cells are of interest, such as for designing handoff schemes. Therefore, such cells are said to have a statistical fractal shape. The comparison of base station deployment \cite{hao2017wireless} for different wireless technology has been summarized in Table \ref{tab:table1}. Since signal received by a user is a function of base station location, strategic placement based on ground surface elevation, location of blockages such as buildings, vegetation etc. improves coverage, expecially at higher carrier frequencies such as at millimeter waves.

\begin{table}[hbtp]
	\begin{center}
		\begin{tabular}{|p{3cm}|p{2.5cm}|p{4cm}|p{4cm}|}
			\hline
			\multicolumn{1}{|c|}{Cellular Network} & \multicolumn{1}{|c|}{3G}  & \multicolumn{1}{|c|}{4G} & \multicolumn{1}{|c|}{5G}\\
			\hline
			Coverage Feature  & Regular hexagon & Irregular polygon & Statistical fractal shape Macrocells and microcells   \\
			\hline
			Deployment &  Macrocells  & Macrocells and microcells & Ultra-dense small cells \\
			\hline
			BS Density & Low & Medium & High \\
			\hline
			Transmission Power of Macro Cell & High & High & High \\
			\hline
			Transmission Power of Small Cell & N/A & N/A & Low \\
			\hline
			Interference & Low & Medium & High \\
			\hline 
			Coverage Redundancy & Low & Medium & High \\
			\hline 
			Wireless Fractal Phenomenon & No & No & Yes \\
			\hline 
		\end{tabular}
		\caption{Comparison of 3G, 4G and 5G.}
		\label{tab:table1}
	\end{center}
\end{table}

Recently, drones have entered the commercial consumer market and the cost of acquiring aerial imagery has also reduced. In addition to RGB images, aerial infrared, depth data and multi-spectral images are also available from them. Due higher resolution in these images compared to satellite images along with recent advances in deep learning, many new applications have emerged such as in surveying \cite{dronesurvey, dronesurvey2, dronesurvey3}, detecting vernal pool \cite{pool}, vehicle counting, precision agriculture \cite{farmbeats, li2016real},  solar panel deployment \cite{sunroof}, power line inspection \cite{martinez2014towards}, wildlife conservation \cite{olivares2015towards}, building inspection \cite{carrio2016ubristes}, land cover segmentation \cite{landcover}, and object segmentation such as buildings, cars, trees etc. The use of these learning techniques has resulted in a drastic reduction in cost and time spent on manual surveying. 

In this thesis, we propose a method leveraging use of location of blockages extracted from aerial drone imagery for finding the optimal mobile base station location. Our approach is expected to be play a vital role in the deployment of base station in emerging mobile communication technologies, including 5G, where higher BS density requirements lead to high deployment costs.

\section{Our Approach}
\begin{figure}[H]
\centering
  \includegraphics[width=0.8\textwidth]{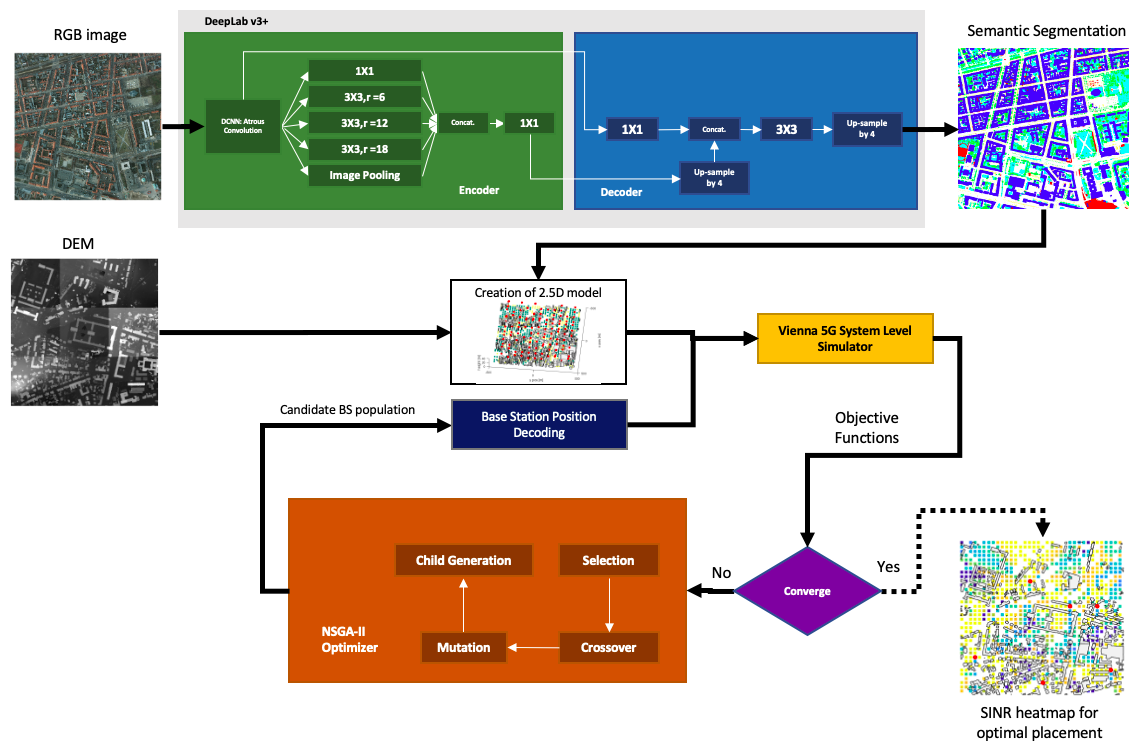}
  \caption{Overview of the proposed pipeline for BS Placement.}
  \label{fig:pipeline}
\end{figure}

A block diagram of our pipeline has been shown in Fig. \ref{fig:pipeline}. In the first part, we perform semantic segmentation to extract location of objects such as buildings, roads, vegetation etc. from aerial drone RGB images.  Then, the segmentation map along with the elevation data (DSM) is used to create a 2.5D model of the scenario. Candidate base stations, user and building locations are also generated based on the segmentation map. 

After this, in the second part, NSGA-II \cite{deb2002fast} is used to select the best base station location based on the SINR of users calculated using Vienna 5G System Level Simulator. This method is inspired by the approach in \cite{gupta2012nsga} for placing cameras for surveillance. 

Although semantic information such as location of buildings, vegetation, elevation of ground surface can be acquired from map service providers such as Google Maps and OpenStreetMaps \cite{haklay2008openstreetmap}, accurate elevation data (height of buildings) is currently not available for buildings, such elevation data is only available for a limited set of locations from these providers as they have started to stitch satellite images with drone images in their solution. Considering the elevation data of the ground surface and buildings for choosing BS location leads to improved coverage analysis especially in the case of higher carrier frequency such as for millimeter waves (MMW). Therefore, we propose the use of aerial drone imagery in our solution.

\section{Organization of Thesis}
The rest of the thesis is organized as follows. Chapter \ref{Chapter2} provides a literature review of all the approaches used by researchers as well as industry practitioners for placement of base stations to improve wireless network performance.  In Chapter \ref{Chapter3}, we describe variuos concepts associated with NSGA-II and its algorithm. Next, we explain variuos deep learning architectures used for semantic segmentation of aerial drone imagery in Chapter \ref{Chapter4}. This is followed by results and analysis in  Chapter \ref{Chapter5} where we find optimal base station placement locations for two areas in Potsdam city along with limitations of our approach. Finally, conclusions and future work has been  presented in Chapter \ref{Chapter6}.

\section{Concluding Remarks}
Our base station placement approach using aerial semantic segmentation would have the benefit of providing superior coverage at a lower deployment cost in a wide variety of areas. Mobile data services such as booking a cab, Wireless Sensor Network (WSN) for monitoring pollution, edge computing devices can particularly benefit from improved coverage. In all cases, the SINR heatmap for optimal base station deployment could act as reference for developing/deploying applications at a lower cost. Our approach is novel as it uses semantic information for determining the optimal location of mobile base stations. This is expected to be superior as the users can expect where there would be sufficient signal strength based on their locations. For instance, the approach can be used to find an optimal location such that all buildings and roads have a minimum SINR.

%% file: Chapters/Chapter2.tex
% Chapter Template

\chapter{Literature Review} % Main chapter title

\label{Chapter2} % Change X to a consecutive number; for referencing this chapter elsewhere, use \ref{ChapterX}

\lhead{Chapter 2. \emph{Literature Review}} % Change X to a consecutive number; this is for the header on each page - perhaps a shortened title

%----------------------------------------------------------------------------------------
%	SECTION 1
%----------------------------------------------------------------------------------------

In this Chapter, we will discuss various approaches that have been used for improving coverage in wireless networks based on location of base stations. Base stations  typically deployed in terrestrial areas have been used to serve mobile users. Recently, aerial base stations, i.e., drones or UAVs with mounted base stations have emerged as a new means of  managing network load and coverage. In this chapter, prior research work on base station placement for both of the above categories have been discussed. For terrestrial base stations, we have covered placement by solving an optimization problem as well as using by using RF planning tools.

\section{Terrestrial Base Station Placement as an Optimization Problem}
\label{sec:litusr}
In this category, researchers have explored base station placement for both indoor and outdoor scenarios. In indoor scenarios, the layout of the building was taken as input from the user. However, the use of building location for base station placement in outdoor scenarios has been limited. The motivation behind this is primarily that in earlier generations of wireless communication techonology, such as 3G, attenuation due to blockages such as buildings was less.

\begin{itemize}
	\item Indoor Scenarios:
Many research groups have explored indoor base station placement by grid search like techniques. Initial works, such as \cite{rappaport2001method, rappaport2006method}, involved the manual placement of base stations in multiple configurations, predicting the coverage using a statistical channel model and selecting the most optimal configuration.  This was then improved in \cite{rappaport2007system}, where the base station configurations were automatically generated exhaustively since the input search space is not large for small indoor building scenarios.  Cutrer \textit{et. al.} \cite{cutrer1997measurement} proposed an approach in which measurements are made corresponding to different base station configuration for selecting the best one. In \cite{butterworth2000base}, Butterworth \textit{et. al.} have  shown how various predetermined placement strategies affect communication performance for in-building scenarios.

The use of optimization algorithms for placement was proposed by Ephremides \textit{et. al.} in \cite{ephremides1999method}. They use methods such as modified steepest descent method, downhill simplex method, Hopfield neural network etc. in their solution. Wright \textit{et. al.} in \cite{wright1998optimization} have optimized the fraction of users inside a building for whom the signal strength is greater than a threshold using Nelder Mead simplex method. In \cite{inasawamethod}, Chiba \textit{et. al.} used Genetic Algorithm for finding the optimal base station configuration inside a building.

	\item Outdoor Scenarios: 
Placement approaches based on evolutionary algorithms have been used in \cite{article, inproceedings, lee2015base} to improve network performance. In \cite{article}, the authors have formulated a weighted average of coverage, interference and financial cost as a single objective function and used Genetic Algorithm for solving it. They also did not consider the effect of blockages. Raisanen et. al. in \cite{inproceedings}, have formulated the base station placement problem as a multiobjective optimization and used Simple Evolutionary Algorithm for Multiobjective Optimization (SEAMO) for solving it. They predefine user locations as Reception Test Points (RTPs), and maximize RTPs covered by cells sites and RTPs with interference less than a threshold. Further, in \cite{lee2015base}, the authors maximize the lowest user throughput normalized by the traffic demand using a custom variant of standard evolutionary algorithm. In \cite{han2001genetic}, the authors have again optimized a weighted average of functions representing coverage and economy of deployment. They have suggested a new real number based representation of base station location for representing a placement configuration as well as custom mutation and cross-over operations in the standard genetic algorithm.

Toros \textit{et.al.} in \cite{torHos2010algorithm} have proposed  an iterative k-means based approach to find the optimal location based on user density. Initially, they choose the center of the deployment area as the location of the first base station. Then, the algorithm is terminated if all users are covered. This is checked using a custom metric defined as the distance of the user from the base station weighted by the demand. In case the users are not covered, k means is used with number of clusters as 2. The 2 centroids are used as the new location of base stations and again it is checked if all users are covered. This process is continued till all users are covered by increasing the number of clusters in every iteration. 
\end{itemize}

\section{Terrestrial Base Station Placement using RF Planning Tools}
\label{sec:litrf}
Despite the placement approaches described above, telecom operators primarily use commercial RF planning tools like WinProp \cite{winprop} and Wireless InSite \cite{wirelessinsite} for deploying their network because the actual deployment sites are constrained by presence of buildings, objects etc. RF engineers select candidate locations and check various configurations of base station for best coverage and deployment cost trade-off. Some examples of  tools for coverage simulation based on ray tracing and statistical channel models have be described below:

\subsection{Ray Tracing Channel Model based Tools}
Some simulators model the wireless network scenario by considering the effect of blockages in a deterministic manner using ray tracing. In this method, given a transmitter and a receiver, all the paths based on direct radiation, reflection and diffraction are computed at the receiver. The signal at the receiver from the multi-paths are accumulated and the channel parameters are then computed. Examples of these tools for coverage simulation include:

\begin{itemize}
	\item WinProp :
	This tool from Altair Feko offers simulation of wireless network scenarios using their proprietary ray tracing model solver and is vastly used by Telecom operators in base station deployment. As an example, the effect of blockages such as buildings on received power for a single base station operating at 28 GHz deployed in the city of Munich is shown in Fig. \ref{fig:winprop}. The simulation was performed in WinProp \cite{winprop} Student Edition. It runs on CPU but is relatively fast as compared to other ray tracing wireless simulators.
	
	\begin{figure}[H]
		\centering
		\includegraphics[width=0.8\textwidth]{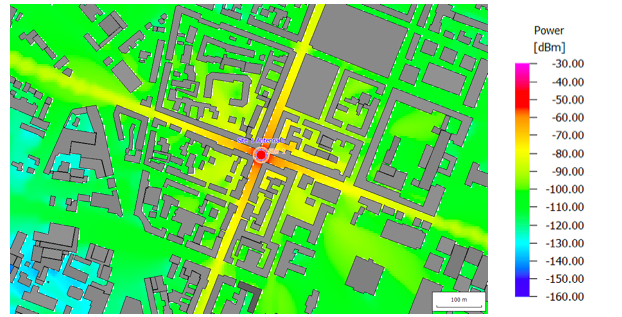}
		\caption{Received Power heatmap from Winprop}
		\label{fig:winprop}
	\end{figure}

	\item RWTH mm-Wave Planning Tool \cite{simic2017coverage}: Developed by Institute for Networked Systems at RWTH Aachen University. The tool is available as a set of open source MATLAB scripts with simulations based on a channel model which runs ray tracing  iteratively on a CPU. It is  relatively slow. For a $700m$ by $700m$ area in Frankfurt, the heatmap of received power from the simulator after one iteration for a single base station operating at 28 GHz has been shown in Fig. \ref{fig:rwth}. It took around 2 minutes to run on a 4 core Intel i7 processor running at 3 GHz.
	
	\begin{figure}[H]
		\centering
		\includegraphics[width=0.7\textwidth]{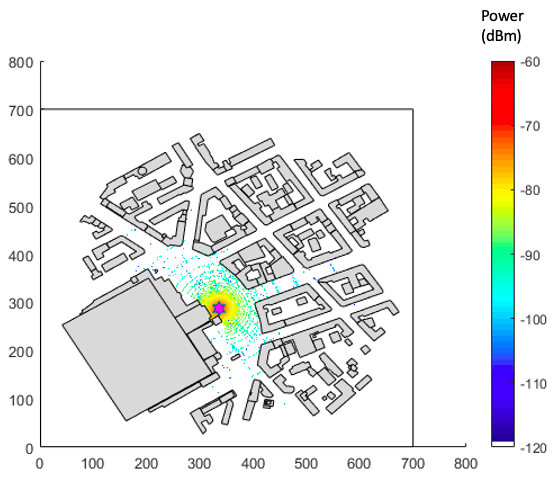}
		\caption{Received Power heatmap from RWTH mm-Wave Planning Tool}
		\label{fig:rwth}
	\end{figure}
	
	\item Huawei 5G Wireless Network Planning Solution \cite{huawei} : Uses a proprietary beam based ray tracing propagation model with GPU support. Apart from direct, reflected and refracted rays, it also supports diffraction after reflection and vice-versa in its simulations. It is only available for commercial use.
	
	\begin{figure}[H]
		\centering
		\includegraphics[width=0.7\textwidth]{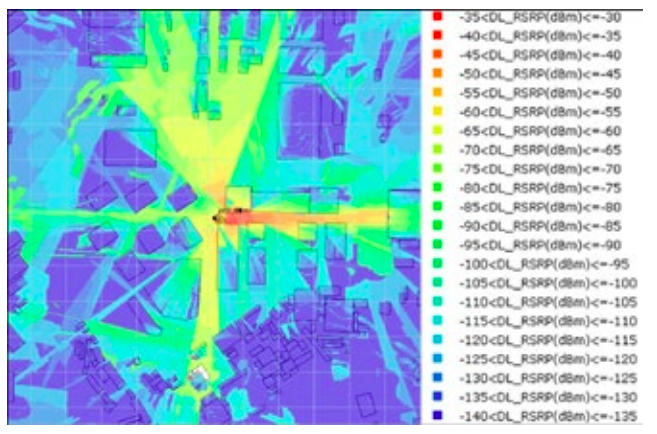}
		\caption{RSRP from Huawei 5G Wireless Network Planning Solution}
		\label{fig:huawei}
	\end{figure}

	Unlike the rest of the simulators, this tool also offers an iterative approach to select base station sites based on the ray tracing simulations. An example heatmap of Reference Signal Received Power (RSRP) simulated from the software has been shown in Fig. \ref{fig:huawei}. This method does not guarantee optimal placement.
\end{itemize}

\subsection{Statistical Channel Model based Tools}
These tools use a statistical channel model based on emperically measured data in their simulators.  Many of these channel models are standardized and are well known. Examples include NYUSIM 5G Channel Model \cite{sun2017nyusim}, WINNER II channel Model \cite{meinila2009winner, zetterberg20052003}, and Rayleigh fading channel model \cite{parsons2000mobile}. Such simulators are relatively faster than simulators based on ray tracing but are also less accurate in predicting the performance of the wireless network. Two of the widely used tools have been listed below:

\begin{itemize}
	\item Vienna 5G System Level Simulator \cite{Vienna5GSLS}: Developed by TU Wien. It is open source under academic license and uses statistical channel models for simulation of coverage on CPU. The tools allows the user to place cuboidal blockages, base stations and users at arbitrary locations and simulate the wireless network. It automatically selects the appropriate channel model based on location of transmitter, receiver, and blockages. Performance can be evaluated in terms of SINR and average user throughput. The current version v1.0, however, does not simulate beam steering and does not have support for millimeter wave channel model.
	\item NYUSIM \cite{sun2017nyusim}: Developed by NYU WIRELESS. The simulator is the first open source tool providing access to channel models at millimeter wave frequencies (28 to 73 GHz) as well. The tool however does not currently have the capability to perform coverage simulations using layout of blockages. The simulator also models the effect of human blockage, foliage loss, humidity, temperature, and barometric pressure which are important for millimeter wave channels. 
\end{itemize}

\section{Aerial Base Stations Placement}
\label{sec:litabs}
With newer generations of wireless communication techonology such as 5G, an approach to compensate for insufficient terrestrial infrastructure is the use of aerial base stations. Their objective is to provide temporal coverage whenever ground base stations are not accessible during hotspots or due to damaged base stations when natural disasters occur. A drawback of aerial base stations is that they can last only as long as their battery backup capacity lasts.

In \cite{inproceedingsaerial1}, the position of base stations is updated in real time to handle network load. The authors have used reinforcement learning to achieve this. They assume that users are moving with a random walk model and constantly evaluate SINR, referred to as Quality of Service (QoS) in the paper. As soon as the QoS drops below a desired threshold, they update the position of the aerial base station based on a state-action knowledge matrix, which is learnt using Q learning. The reward function $r_t$ and knowledge matrix $Q(s_t,a_t)$ update equations have been shown below:

Reward function  : $r_t = QoS_t - QoS_{t-1}$
 
Knowledge matrix update: $Q(s_t,a_t)=\alpha[r_t+\gamma \max Q(s_{t+1},a) - Q(s_t,a_t)]$

At time $t$, the agent state is $s_t$ with $a_t$ being the corresponding action. The learning rate is $0 < \alpha < 1$ and the discount factor is $0 < \gamma < 1$. For exploration and exploitation in the learning process, they use the $\epsilon$-greedy strategy.

Another group of researchers in \cite{inproceedingsaerial2}, have considered the location of users to be fixed. They model the aerial base station placement problem as a optimization problem in which they maximize the number users served by the drone cell with constraints on its location for ensuring a minimum signal strength according to the path loss model. This optimization problem was solved using bisection search algorithm.

\section{Concluding Remarks}
This Chapter discussed approaches for base station placement which were categorized into terrestrial and aerial. Compared to the works available in literature, our approach primarily deals with base station placement in outdoor scenarios. We formulate the problem as multi-objective optimization by simulating the scenario with blockage locations extracted with a deep learning model. The following chapters present our approach in detail.

%% file: Chapters/Chapter3.tex
% Chapter Template

\chapter{Non-dominated Sorting Genetic Algorithm} % Main chapter title

\label{Chapter3} % Change X to a consecutive number; for referencing this chapter elsewhere, use \ref{ChapterX}

\lhead{Chapter 3. \emph{Non-dominated Sorting Genetic Algorithm}} % Change X to a consecutive number; this is for the header on each page - perhaps a shortened title

%----------------------------------------------------------------------------------------
%	SECTION 1
%----------------------------------------------------------------------------------------
\section{Introduction}
\label{sec:nsgaint} 
Evolutionary algorithms refer to the class of optimization algorithms which use a genetic population based metahueristic. It includes Genetic Algorithms (GA), Genetic Programming, Evolutionary Programming, Differential Evolution etc. In Genetic Algorithms, the solution of the optimization problem being solved can be represented by a string of numbers in which the input is randomly perturbed in every iteration for finding the optimal solution. In multi-objective optimization problems, Non-dominated Sorting Genetic Algorithm (NSGA-II) \cite{deb2002fast} is one of the state of the art optimization methods. We choose this method since the placement function is multi-objective in nature and the input space for searching potential base station locations is high-dimensional. This chapter discusses key concepts that would be needed to understand this algorithm.

\subsection{Dominance and Pareto-optimal Front}
\label{sec:nsgadom}
\begin{figure}[H]
	\centering
	\includegraphics[width=0.75\textwidth]{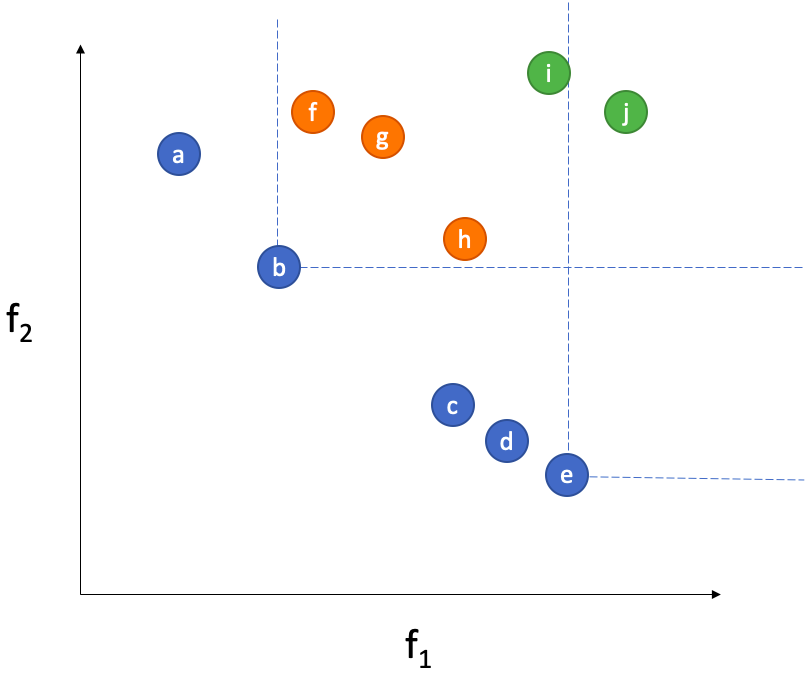}
	\caption{Dominance and pareto-optimal front}
	\label{fig:pareto}
\end{figure}

In a multi-objective setting, we have a set of objectives, each of which needs to be maximized or minimized. The optimal solution is dependent on each of the individual objective values and the concept of non-dominance is used to address this. A solution is called non-dominated / pareto-optimal, if none of the objective functions can be improved in value without degrading some of the other objective values. For instance, in Fig. \ref{fig:pareto}, two objectives $f_1$ and $f_2$ have been shown on the X and Y axis. Each point represents the value of the two objectives for different solutions. We consider the case of minimizing both the objectives. $b$, is said to dominate $f,g,h,i,j$ as it has a lower value of both objectives. Similarly, $e$ dominates only $j$. 

Pareto-optimal front is used to categorize solutions which have a trade-off in terms of optimal value of different objectives. No solution in the pareto-optimal front dominates the other solutions present in it. For example, let $n_p$ represent the number of solutions which a solution dominates. Then the set of solutions which have $n_p=0$ is called the pareto-optimal front. The blue colored dots represent front $F_1$, orange represents front $F_2$ and green represents  front $F_3$.

\subsection{Non-dominated Sorting/Pareto-dominance}
In non-dominated sorting, multiple solutions of a generation have to be grouped into fronts, as explained above, and sorted in decreasing order of optimality. Let $p$ be a solution of the current generation $P$. As earlier, $n_p$ represents the number of solutions which solution $p$ dominates and $S_p$ denote the set of those solutions. 

Initially, $n_p$ and $S_p$ is calculated for each solution. All solutions with $n_p = 0$ can directly be assigned to the pareto-optimal front $F_1$.

\begin{algorithm}[H]
	\For{\upshape each p in P}{
		$S_p = \phi$\\
		$n_p=0$\\
		\For{\upshape each q in P}{
		{\uIf{\upshape p dominates q}
			{Add q to the set of solutions dominated by p, i.e., $S_p$}
			\Else{Increment $n_p$.}
		}
	}
		\If{$n_p=0$}{
			$p_{\text{rank}}=1$\\
			$F_1=F_1 \cup {p}$
		}
	}
\caption{Assignment of set of dominating solutions}
\end{algorithm}

Next, for each solution $p$ in the first front $F_1$,  every member $q$ of its set $S_p$ is visited and its count ($n_q$) is reduced by 1. In doing so, if for any solution $q$, $n_q = 0$, then it is put in a new list $Q$ corresponding to the second non-dominated front. This process is continued till no solution is assigned to the current front.

\begin{algorithm}[H]
		$j=1$\\
		 \While{$F_j \neq \phi$}{
			$Q = \phi$\\
			\For{$p \in F_j$}{
				\For{$q \in S_p$}{
					$n_q = n_q -1$\\
					\If{$n_q = 0$}{
						$q_{\text{rank}} =  j +1$\\
						$Q = Q \cup {q}$
					}
				}
			}
		$j = j+1$\\
		$F_j = Q$
		}
	\caption{Assignment of front to solutions}
\end{algorithm}

\subsection{Crowding Distance}

\begin{figure}[H]
	\centering
	\includegraphics[width=0.75\textwidth]{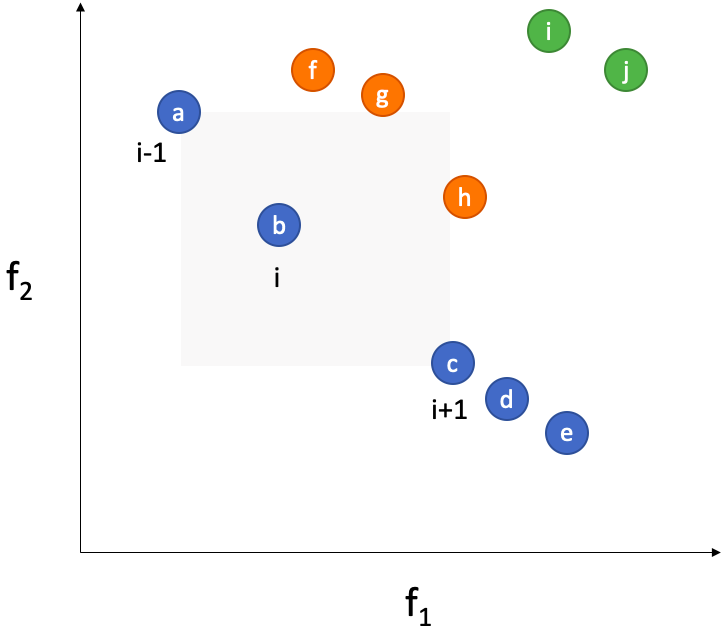}
	\caption{Calculation of crowding distance}
	\label{fig:crowd}
\end{figure}

In order to promote exploration of new solutions in less dense areas of the solutions generated in previous steps, the crowding distance metric is defined. A solution with higher crowding distance denotes a lower density of solutions around it. It is used to sort solutions within a front. For example in Fig. \ref{fig:crowd}, the crowding distance of solution $b$ will depend on solutions $a$ and $c$. $b$ will also have a higher crowding distance than solution $d$. The procedure to compute crowding distance is described below:

\begin{algorithm}
	l=$|I|$\\
	\For{\upshape each i,}{Set I[i]=0}
	\For{\upshape each objective m}{
		$I_m = \text{sort(I,m)}$\\
		$I[1]_{\text{distance}}=I[l-1]_{\text{distance}}=\infty$\\
		\For{\upshape i=2 to l-1}{
			$I[i]_{\text{distance}}=I[i]_{\text{distance}}+(I[i+1]_m-I[i-1]_m)/(f_m^{\text{max}}-f_m^{\text{min}})$\\
		}	 
	}
	\caption{Assignment of crowding distance}
	\label{algo:nsgaPart}
\end{algorithm}

For every objective function, the solutions in the front $I$ are sorted in increasing order of crowding distance. The boundary solutions are assigned an infinite distance and the intermediate solutions are assigned a distance 
equal to the absolute normalized difference in the function values of two adjacent solutions. The steps of the algorithm have been formally described in \ref{algo:nsgaPart}. Crowding distance is stored in $I_{\text{distance}}$, the maximum and minimum value of each objective $m$ is stored in $f_m^{\text{max}}$ and $f_m^{\text{min}}$.

\subsection{Tournament Selection}
It is the process of selecting chromosomes from a population for generating child chromosomes. Individuals from a population are chosen at random and "tournaments" ,ie, comparison based on non-dominated sorting are conducted for selection. 

\subsection{Binary Two Point Crossover}
In this step, given two parent chromosomes, two new child chromosome are generated based on a crossover probability ($p_c$). 

\begin{figure}[H]
	\centering
	\includegraphics[width=1\textwidth]{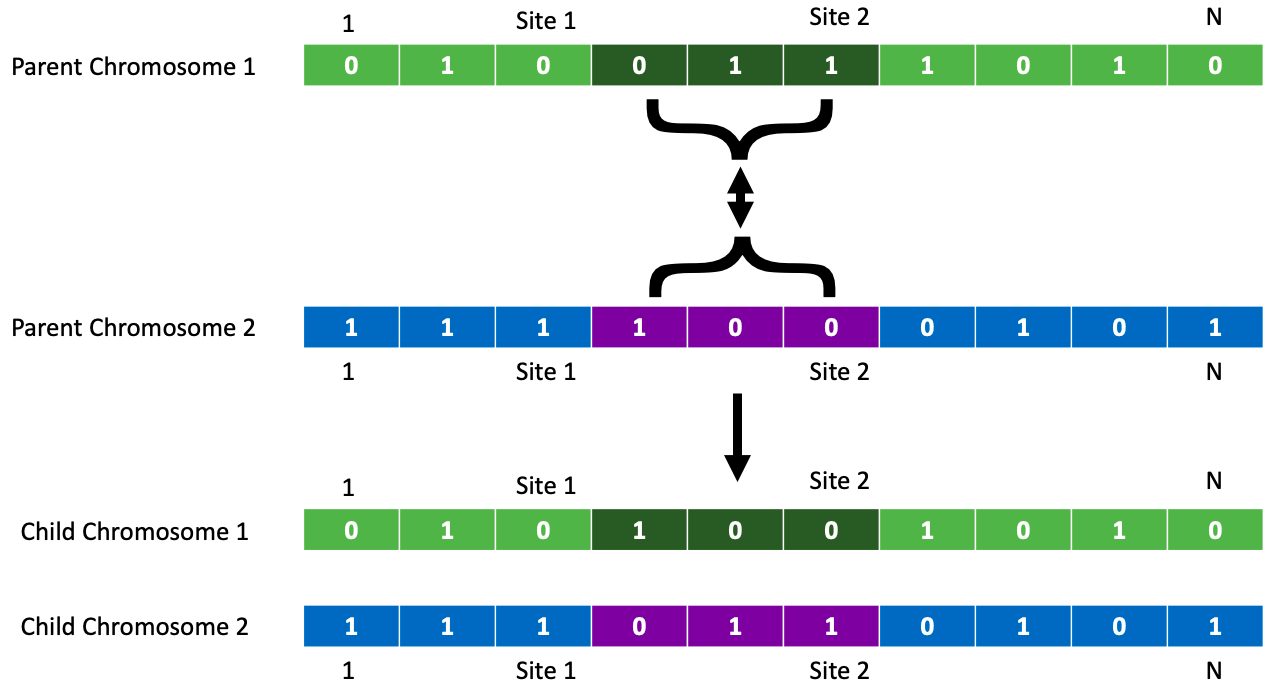}
	\caption{Binary Two Point Crossover}
	\label{fig:crossover}
\end{figure}

As shown in Fig. \ref{fig:crossover}, two indices (sites) are randomly selected and the bits in that range (Site1 + 1 to Site 2) are exchanged between the two parent chromosomes with prabability $p_c$, thereby creating two new child chromosomes. The motivation behind this step is to combine the best genes from parent chromosomes.

\subsection{Mutation}
In mutation, each bit of the binary representation of a chromosome is randomly flipped with a mutation probability $p_m$. This step is performed to ensure genetic diversity in the newly generated population.

\subsection{Algorithm}
Initially, a set of $k$ solutions of optimization problem is generated randomly in the form of a vector of integers, real or binary numbers. Each solution is referred to as a chromosome and the set is known as the parent population $P_t$. The solutions are then evaluated and sorted based on pareto-dominance sort. A child population $Q_t$ is then generated from the parent population through binary tournament selection followed by recombination and mutation. 

The two populations are combined and again assigned fronts using pareto-dominance sort and then sorted within fronts using crowding distance. Parent population for the next generation $P_{t+1}$ are then chosen as the top $k $ chromosomes from this sorted population. This process is then repeated for a fixed number of generations $\text{nGen}$.

\begin{figure}[H]
	\centering
	\includegraphics[width=1\textwidth]{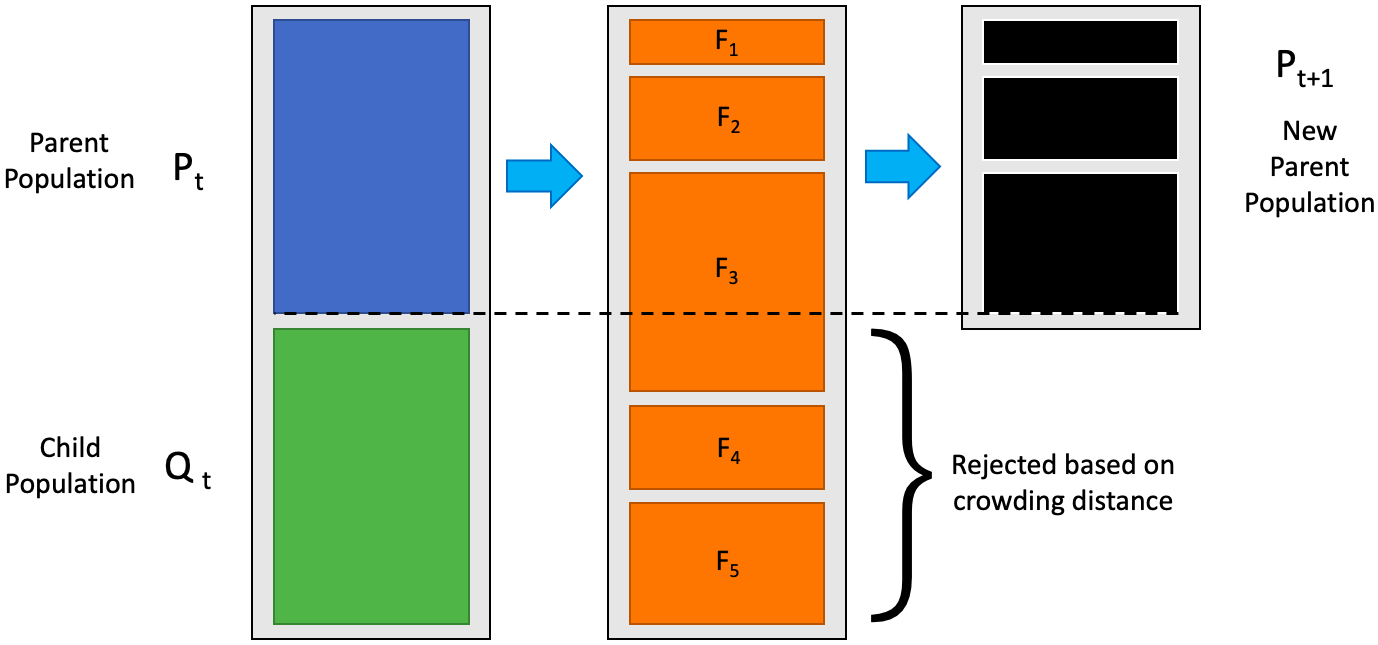}
	\caption{Population generation in NSGA-II}
	\label{fig:nsga_gen}
\end{figure}

\begin{algorithm}[H]
	\textbf{Parameters: } pSize, nGen, pCross, pMut\\
	\textbf{Initialize:} Randomly generate the parent population\\
	Set generation count $g  = 0$\\
	Compute values of Objective Functions\\
	Perform Pareto-dominance Sort and assign rank\\
	Generation of Child Population\\
		\quad{Perform Binary Tournament Selection}\\
		\quad{Perform Crossover followed by Mutation}\\
	\For{\upshape $g \gets 1$ to nGen}{
		\For{\upshape each chromosome in child population}
		{Use Pareto - dominance Sort to assign rank.\\
		Assign crowding distance to each front and sort.\\
	     }
     Creation of new generation\\
     	\quad{Perform Binary Tournament Selection}\\
     	\quad{Perform Crossover followed by Mutation}\\
	}
	\caption{Non-dominated Sorting Genetic Algorithm}
\end{algorithm}

\section{Modelling Base Station Placement as a Multi-objective Problem}
\label{sec:nsgamod}
Since the objective function values are to be calculated for each chromosome in the population, we needed an open source simulator which simulated the wireless network in minimal time. Therefore, we used Vienna 5G simulator \cite{Vienna5GSLS}. 

\begin{figure}[H]
  \centering
     \includegraphics[width=0.75\textwidth]{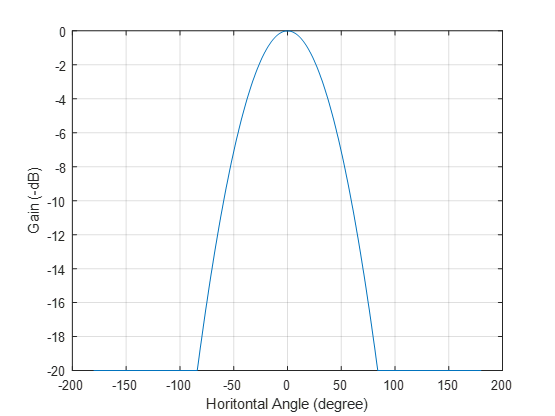}
  \caption{Radiation pattern of antenna used in simulation.}
  \label{fig:a_pattern}
\end{figure}

The simulator is used in lite mode since we are only interested in coverage optimization. To simulate Long Term Evolution (LTE) scenario, we used a carrier frequency of 2 GHz with each three sector base station composed of 3 antennas having a Half Power Beam Width (HPBW) of 65 degrees \cite{3gpp.36.942}. The radiation pattern is shown in Fig. \ref{fig:a_pattern}. 

Each base station location $(x,y)$ is binary encoded and an array of binary encoded locations is used to represent a particular base station configuration or a chromosome in NSGA-II. 

\subsection{Use of Semantic Segmentation}
Each of the segments corresponding to buildings were used along with the elevation data from the DSM to create blockages. Users were then placed at an elevation of $2$m above the elevation of segments corresponding to \textit{impervious surfaces}, \textit{low vegetation} and \textit{clutter}. Candidate base station locations were created on a regular grid spanning the entire area except where \textit{trees} , \textit{clutter} or \textit{car} segments were present. This was done since the \textit{clutter} class contained water bodies and other objects such as containers, tennis courts, swimming pools etc., which could not be considered as candidate locations. 

\subsection{Objective Functions}
\label{sec:objfun}
We now formally introduce the multi-objective function used for NSGA-II in our approach. Let each $SINR_i$ denote the signal strength of the $i^{th}$ user, $N$ be the total number of users, $M$ be the number of BS deployed, $M_{\text{max}}$ be the maximum number of BS we wish to deploy in a given area, and $SINR_{\text{threshold}}$ be the threshold above which we want to maximize the signal strength. The following three objective functions are optimized in our approach:
\begin{itemize}
    \item  Maximize the total SINR of users near buildings and roads: 
    \begin{gather}
       \min \left(-\sum_{k} SINR_k \right)
       \label{eq:obj1} 
    \end{gather} 
    Here, $k$ denotes the users near buildings and roads.
    
    \item  Minimize the number of BS deployed :
    \begin{gather}
        \min M \\
        \text{subject to} \nonumber \\
        1 \leq M \leq M_{\text{max}}
        \label{eq:obj2} 
    \end{gather}
    
    \item Maximize the number of users with SINR greater than a threshold :
    \begin{gather}
        \min \left(-\sum_{i=1}^{N} \mathbbm{I}{ (SINR_i>SINR_{\text{threshold}}) } \right)
        \label{eq:obj3} 
    \end{gather}
    Here, $\mathbbm{I}$ denotes the indicator function.
\end{itemize}

\section{Modifications to Vienna 5G System Level Simulator}

The simulator was also modified to support arbitrary shaped blockages. The following two changes were made for this:
\begin{itemize}
	\item A class PolygonBuilding was defined which contains a wall with $n$ vertices corresponding to the roof and $n$ walls with $4$ vertices corresponding to the side walls. Cities were represented using an array of PolygonBuilding objects. As an example, the 2.5D model of an area in Potsdam from the ISPRS dataset \cite{potsdam} has been shown in Fig. \ref{fig:3dscenario}. 
	
	\begin{figure}[H]
		\centering
		\includegraphics[width=1\textwidth]{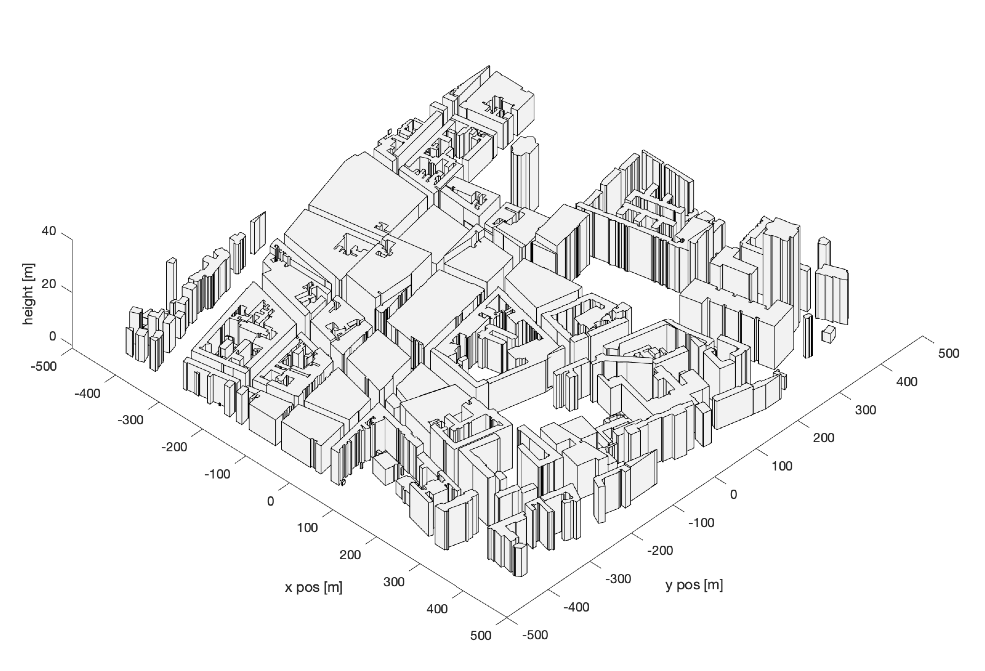}
		\caption{Potsdam Urban Scenario with arbitrary shaped buildings.}
		\label{fig:3dscenario}
	\end{figure}

	\item In order to check line of sight between transmitter and receiver, the ChunkSimulation class was modified. The function to check line of sight was altered using the MATLAB geometry toolbox for 2D/3D geometric computing \cite{matGeom}.
	
\end{itemize}
 
\section{Concluding Remarks}
In this chapter, we discussed various concepts such as non-dominated sorting, pareto-optimal front, crowding distance, tournament selection,  binary two point crossover, mutation and explained the working of NSGA-II. This was followed by explanation of modelling base station placement as a multi-objective problem and finally, various modifiations required to be done in Vienna 5G System Level simulator were discussed. 

%% file: Chapters/Chapter4.tex
% Chapter Template
%cite example uses of segmentation in beginning
\chapter{Semantic Segmentation} % Main chapter title

\label{Chapter4} % Change X to a consecutive number; for referencing this chapter elsewhere, use \ref{ChapterX}

\lhead{Chapter 4. \emph{Semantic Segmentation}} % Change X to a consecutive number; this is for the header on each page - perhaps a shortened title

%----------------------------------------------------------------------------------------
%	SECTION 1
%----------------------------------------------------------------------------------------
\section{Introduction}
\label{sec:ssint}
\begin{figure}[H]
	\centering
	\includegraphics[scale=0.4]{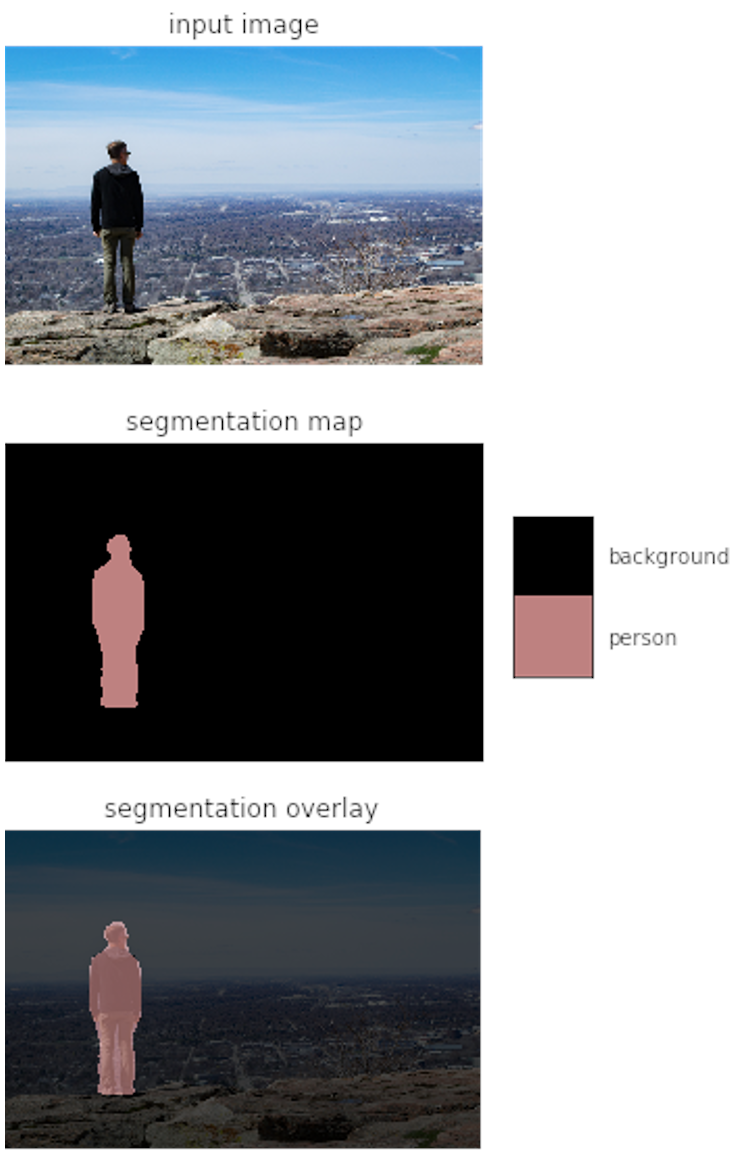}
	\caption{Example of semantic segmentation of images.}
	\label{fig:seg_ex}
\end{figure}

Image segmentation refers to the task of grouping together pixels that are homogeneous with respect to characteristics such as color, texture, etc. When the characteristic has a semantic meaning, such as an object, the task is referred to as semantic segmentation. Images are used as input to create regions representing different objects. Supervised learning models such as Convolutional Neural Networks (CNN) have primarily been used for this task. CNNs learn a powerful hierarchy of visual features. An example has been shown in Fig. \ref{fig:seg_ex}.

Prior to the bloom of deep learning, this task was performed by using filter bank on the input image followed by clustering and assigning every pixel labels using conditional random field. Such methods have been outperformed by approaches based on deep learning which use an encoder-decoder style CNN for performing the task. The network performs classification for each pixel. Output segmentation map of objects is useful in determining object boundaries and has found applications in autonomous driving \cite{nvidiaDriving}, robot vision \cite{yeboah2018semantic}, tumour detection \cite{kamnitsas2017ensembles} etc.

In this chapter, we will primarily explain common concepts used in semantic segmentation using deep neural networks. Then we will discuss DeepLabv3+, a state of the art neural network for semantic segmentation. This is followed by explanation of using the segmentation map and DSM to create a 2.5D model.

\subsection{Atrous Convolution}
\begin{figure}[H]
	\centering
	\includegraphics[width=0.35\textwidth]{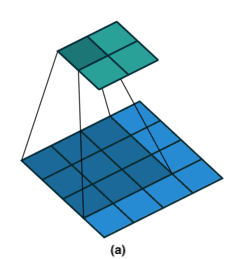}
	\includegraphics[width=0.35\textwidth]{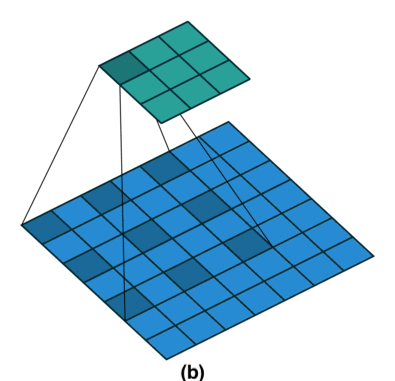}\\
	\includegraphics[width=0.15\textwidth]{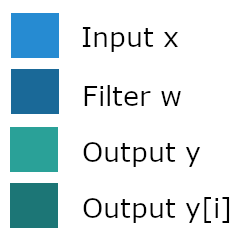}
	\caption{Convolution and Atrous Convolution}
	\label{fig:aconv}
\end{figure}

Atrous convolution (also known as dilated convolution) operator for 2D signals $x$ (input) and $w$ (filter) indexed with $i$  is defined as :
\begin{gather}
y[i] = \sum_{k}x[i+r.k]w[k]
\end{gather}
Here, $r$ represents the atrous rate. It signifies the stride for sampling input $x$. When $r=1$, atrous convolution is same as standard convolution. The comparison of the two \cite{dumoulin2016guide} has been shown in Fig. \ref{fig:aconv}. Both the examples are for no zero padding and rate $r=2$ for atrous convolution. $r-1$ elements of input are skipped between consecutive elements of filter weight $w$ in atrous convolution. The motivation behind this step is to increase the receptive field of the filter. Every element of the output feature map for a $3 \times 3$ filter in atrous convolution interacts with a $5 \times 5$ region of input signal because of dilation unlike a $3 \times 3$ region in standard convolution.

The total number of operations for convolution of a $n \times n \times d$ input with $h$ filters of size $k \times k \times d$, producing an output of size $m \times m \times h$ will be:\\
\begin{gather}
	(k \times k \times d) \times (m \times m \times h) = d m^2 k^2h
\end{gather}

One drawback \cite{chen2018encoder} of using atrous convolution instead of standard convolution in a network like ResNet \cite{he2016deep} is that the output stride (defined as the ratio of the input image resolution to the encoder output resolution) cannot be low. For example, for output stride of 16, the feature map from the last 3 residual blocks need to be dilated. Similarly, for output stride of 8, the feature map from the last 26 residual blocks need to be dilated.

\subsection{Atrous Separable Convolution}
Atrous separable convolution involves atrous depthwise convolution with $d$ filters of size $k \times k \times 1$  followed by pointwise convolution with $h$ filters of size $1 \times 1 \times d$. This enables faster computation and larger receptive field as compared to standard convolution with $h$ filters of size $k \times k \times d$.  Atrous depthwise and pointwise convolution have been explained below:

\begin{enumerate}
	\item Atrous Depthwise convolution: In depthwise convolution, as shown in Fig. \ref{fig:depthwise} (a), every channel of an input of size $n \times n \times d$ is convolved with a $k \times k \times 1$ sized filter to get output of size $m \times m \times d$. The number of channels in input and output remain the same, only the height and width changes. For atrous depthwise convolution, the only difference is that standard convolution is replaced with atrous convolution with some rate $r$. In Fig. \ref{fig:depthwise} (b), atrous depthwise convolution with rate $r=2$ has been shown. \\
	The total number of operations in depthwise convolution is :
	\begin{gather}
	(k \times k \times 1) \times (m \times m \times d)  
	= k^2 m^2 d
	\end{gather}
	The use of atrous convolution ensures a larger receptive field for the filter.
	\begin{figure}[H]
		\centering
		\includegraphics[width=1\textwidth]{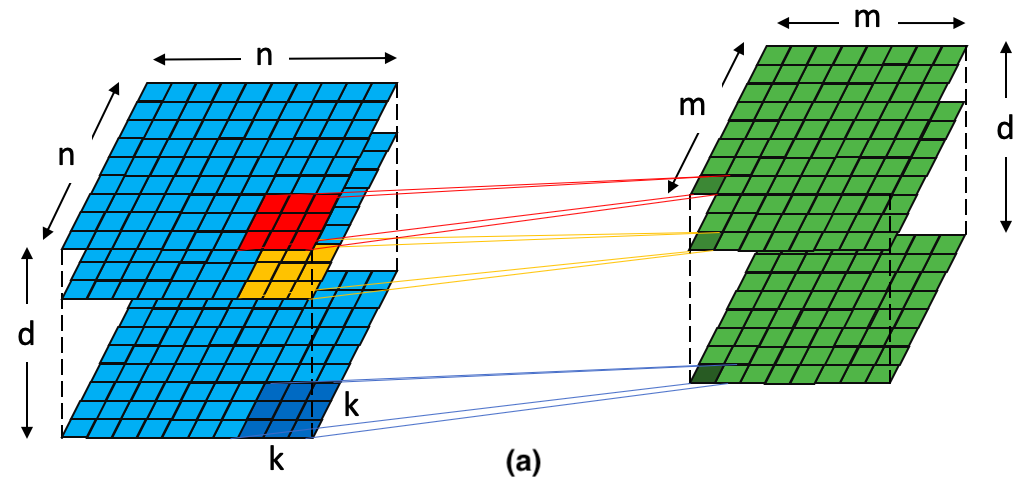}
		\includegraphics[width=1\textwidth]{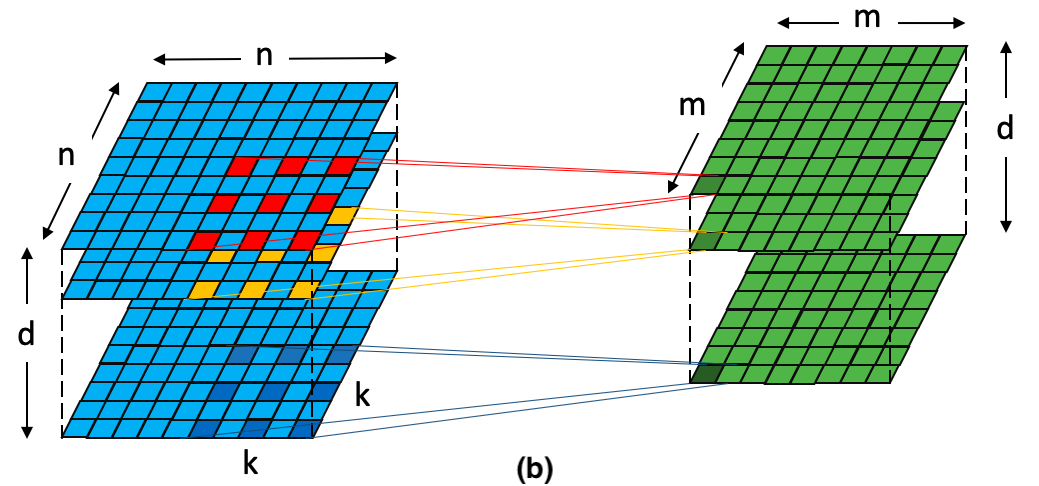}
		\caption{Depthwise Convolution and Depthwise Atrous Convolution}
		\label{fig:depthwise}
	\end{figure}

	\item Pointwise convolution: In pointwise convolution, as shown in Fig. \ref{fig:pointwise}, every visual word of an input of size $m \times m \times d$ is convolved with $h$ filters of size $1 \times 1 \times d$ to get output of size $m \times m \times h$. The number of channels in input and output vary while the height and width remains the same. Only one filter of size $1 \times 1 \times d$ has been shown in the figure for the purpose of illustration.\\
	The total number of operations in pointwise convolution is :
	\begin{gather}
	(1  \times  1  \times  d)  \times  (m  \times  m  \times  h)
	= d m^2 h
	\end{gather}
	\begin{figure}[H]
		\centering
		\includegraphics[width=0.7\textwidth]{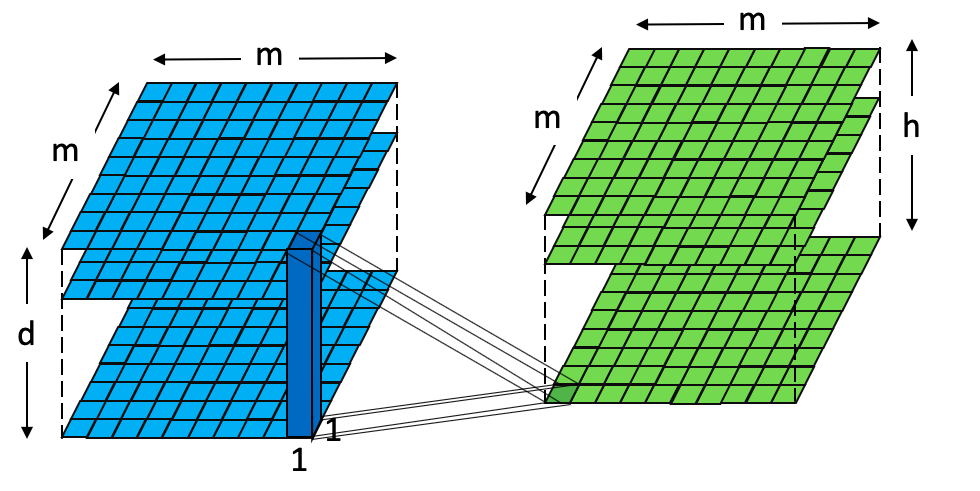}
		\caption{Pointwise Convolution}
		\label{fig:pointwise}
	\end{figure}
	
\end{enumerate}

Therefore, the total number of operations in atrous separable convolution is (atrous separable convolution followed by depthwise separable convolution is) : \\
\begin{gather}
	k^2 m^2 d + d m^2 h = d m^2 (h+k^2)
\end{gather}
We can see that this is less than $d m^2 k^2h$, that is, the number of operations for standard convolution, thereby making improving the computation speed.

\section{Encoder Decoder Structure}
\label{sec:ssfcn}
\begin{figure}[H]
	\centering
	\includegraphics[scale=0.45]{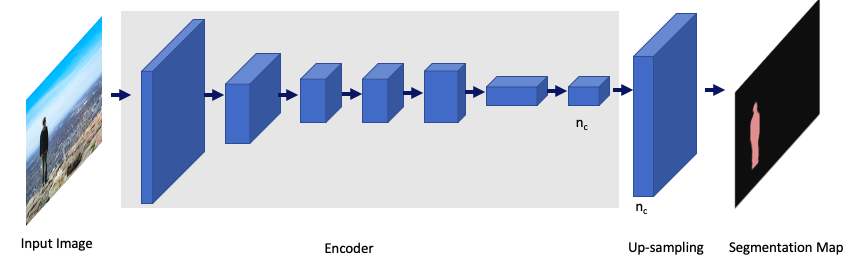}
	\caption{Fully Convolutional Network}
	\label{fig:fcn}
\end{figure}

Convolutional Neural Networks used for image classification typically consists of an encoder module whose output is flattened and passed through fully connected layers for performing classification. An output stride of 32 is typically used. However, semantic segmentation requires pixelwise classification of input image. Therefore, for a denser feature extraction, a lower output stride (8 or 16) is used. Fully Convolutional Networks (FCN) \cite{long2015fully}, use deconvolution to upsample the encoder output and assign labels to each pixel. The representative network architecture has been shown in Fig. \ref{fig:fcn}. Here $n_c$ represent number of channels which are same as the number of classes for which pixel-wise classification is being performed.

\begin{figure}[H]
	\centering
	\includegraphics[scale=0.3]{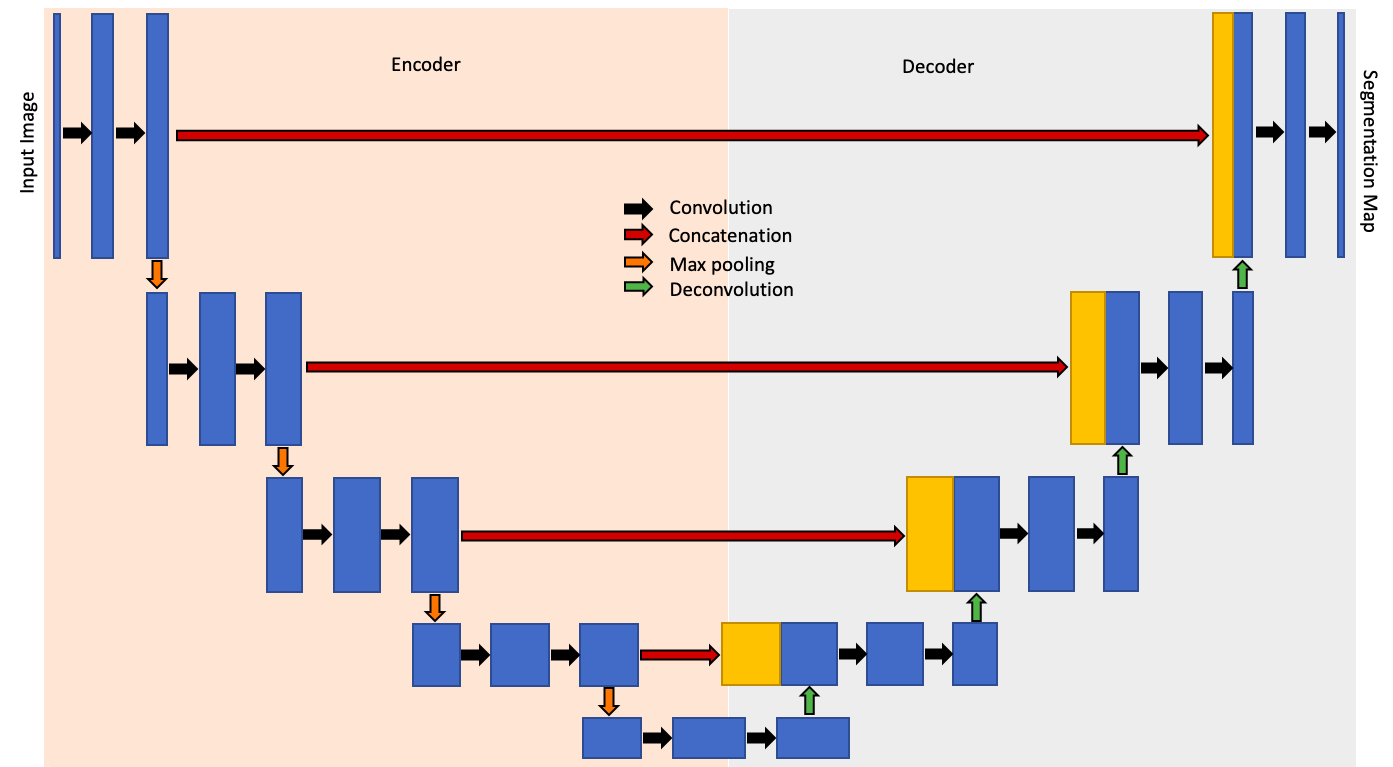}
	\caption{U-Net Architecture}
	\label{fig:unet}
\end{figure}

UNet \cite{ronneberger2015u} improved this architecture by having a deeper decoder and by introducing skip connections between encoder and decoder. The network architecture has been shown in Fig. \ref{fig:unet}. The semantic information increase as we move deeper in the encoder and the segmentation boundaries become finer as we move deeper in the decoder.

\section{Pyramid Pooling}
In this section, we explain the concept of using multiple pooling/strided convolution operations as a means of capturing multi-resolution contextual information. For ResNet, Zhou et. al. \cite{zhou2014object} emperically showed that the receptive field of CNN is smaller than the input image, thereby preventing the network from incorporating the global scenery prior. This issue can be solved by using pyramid pooling to capture multiple receptive fields.

\subsection{Spatial Pyramid Pooling}
\begin{figure}[H]
	\centering
	\includegraphics[scale=0.30]{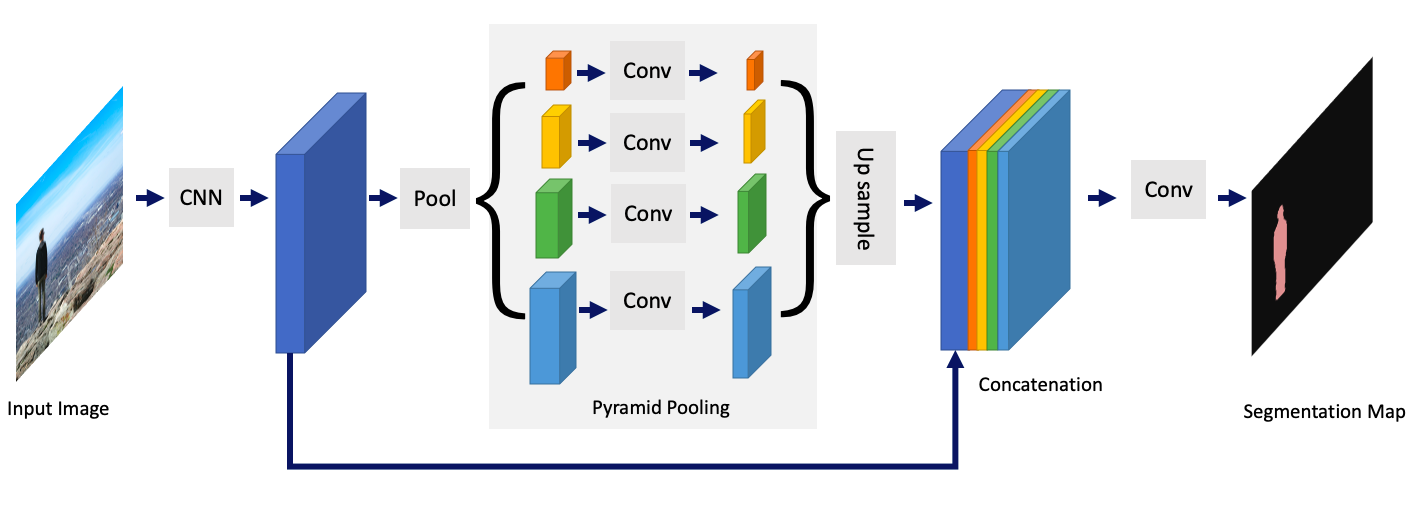}
	\caption{Pyramid Pooling in PSPNet}
	\label{fig:psp}
\end{figure}

Zhao et. al. in \cite{zhao2017pyramid} introduced the concept of spatial pyramid pooling for semantic segmentation in Pyramid Scene Parsing (PSPNet) network. The feature map of the encoder is pooled at multiple grid scales, upsampled to the same height and width, and concatenated as show in Fig. \ref{fig:psp}. 

FCN are less capable to learn contextual relationships, for instance, the probability of an aeroplane being near the sky is more than being on road. They also cause discontinuous segmentation for very small or very large objects. These are attributed to FCN's inability to learn contextual relationship and global information from the receptive field of CNN. 

Spatial Pyramid Pooling addresses the above limitations as it pools or convolves the encoder response map with different strides (resulting in different receptive fields) and then learns mutiscale features from the concatenated output. 

\subsection{Atrous Spatial Pyramid Pooling (ASPP)}
In contrast to SPP, atrous SPP performs atrous convolution with different rates instead of convolving with different strides or pooling. This results in denser feature maps which contains detailed information about object boundaries.

\section{DeepLabv3+}
\label{sec:ssdeeplab}
\begin{figure}[H]
	\centering
	\includegraphics[width=1\textwidth]{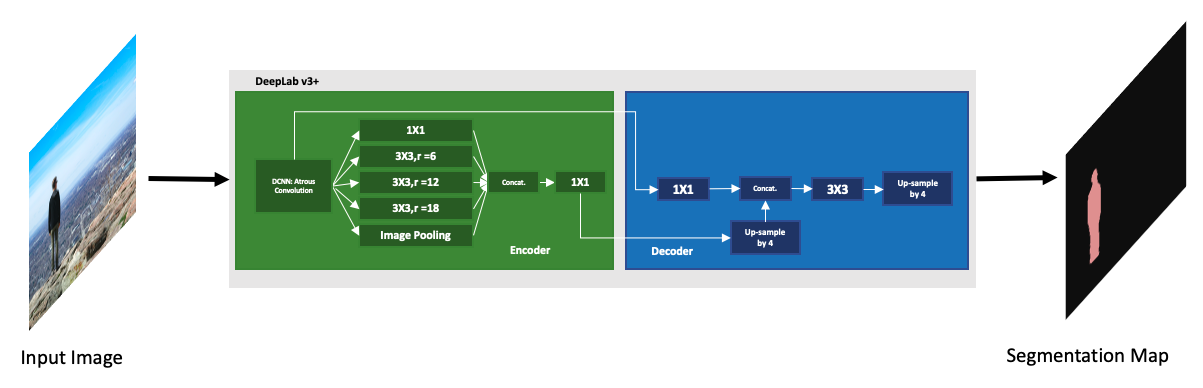}
	\caption{Network Architecture of DeepLabv3+}
	\label{fig:deeplab}
\end{figure}

Chen et. al. in \cite{chen2018encoder}, combined the concepts of atrous spatial pyramid pooling and encoder decoder structure for semantic segmentation. Low level features from the encoder having the same resolution as the upsampled output of the encoder are also concatenated and fed into the decoder. The overall network architecture has been shown in Fig. \ref{fig:deeplab}.

\begin{figure}[H]
	\centering
	\includegraphics[width=0.58\textwidth]{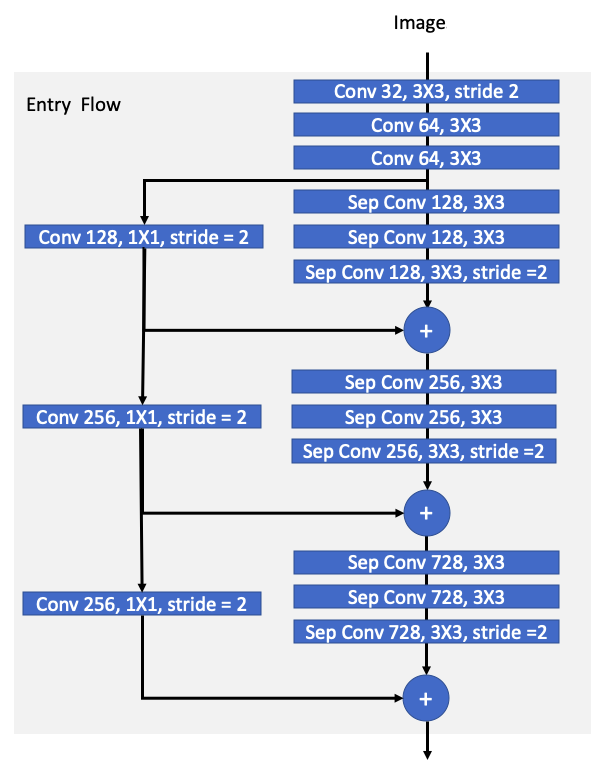}
	\includegraphics[width=0.58\textwidth]{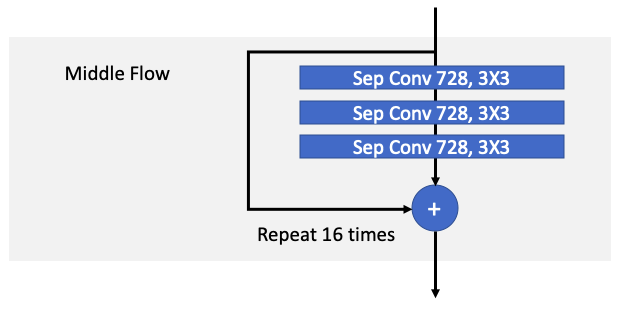}
	\includegraphics[width=0.58\textwidth]{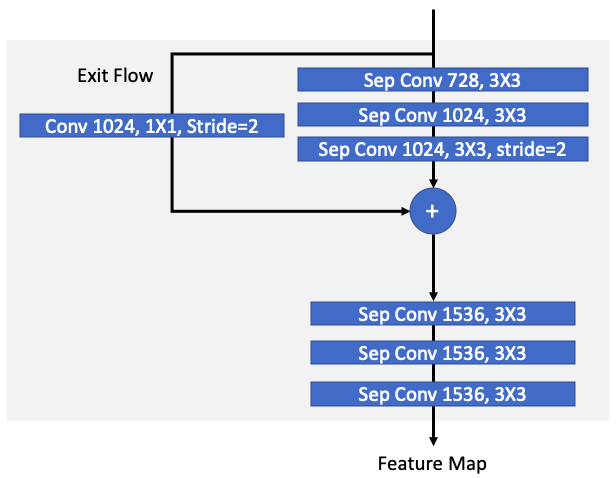}
	\caption{Modified Xception Network}
	\label{fig:xception}
\end{figure}

The Deep Convolutional Neural Network (DCNN) in the encoder shown in Fig. \ref{fig:deeplab} represents the Xception \cite{chollet2017xception} network shown in Fig. \ref{fig:xception}. It is used to extract feature map from the input image. The authors changed the pooling layer in Xception with atrous separable convolution for improved computation speed. The modified network architecture of Xception network has been shown below in Fig. \ref{fig:xception}.

\section{Evaluation Metrics}
\label{sec:ssem}
Evaluation of semantic segmentation has been studied in detail in \cite{csurka2013good}. Since it is essentially a pixelwise classification problem, the metrics used in classification are extended and modified for evaluation in this task. The commonly used terms for evaluating classification are :

\begin{enumerate}
	\item True Positive (TP) refers to the outcome when the model predicts the positive class correctly.
	\item True Negative (TN) refers to the outcome when the model predicts the negative class correctly.
	\item False Positive (FP) refers to the outcome when the model predicts the positive class incorrectly.
	\item False Negative (FN) refers to the outcome when the model predicts the negative class incorrectly.
\end{enumerate}

Further, precision and recall are defined using the above as:
\begin{gather}
\text{Precision} = \frac{\text{TP}}{\text{TP+FP}}\\
\text{Recall} = \frac{\text{TP}}{\text{TP+FN}}
\end{gather}
The extension of these for semantic segmentation have been discussed below.

%\subsection{Evaluation Metrics}
%\label{sec:ssem}
\begin{itemize}
	\item{Class Wise Recall : }
	For each class, it represents the ratio of number of pixels correctly classified to the total number of pixels that class (based on ground truth). A drawback of class wise recall is that a higher recall does not always imply a superior segmentation. For instance, if all pixels are predicted to belong to class A, then recall for class A will be high, but recall for other classes will be low.
	
	\item{Intersection over Union (IoU) : }
	It is the ratio of the area corresponding to intersection of the predicted and ground truth segment to their area of union. As shown in Fig. \ref{fig:iou}, the ground truth segment is the red mask and the predicted segment is the blue mask. It is also known as Jaccard similarity coefficient.
	
	\begin{figure}[H]
		\centering
		\includegraphics[width=0.75\textwidth]{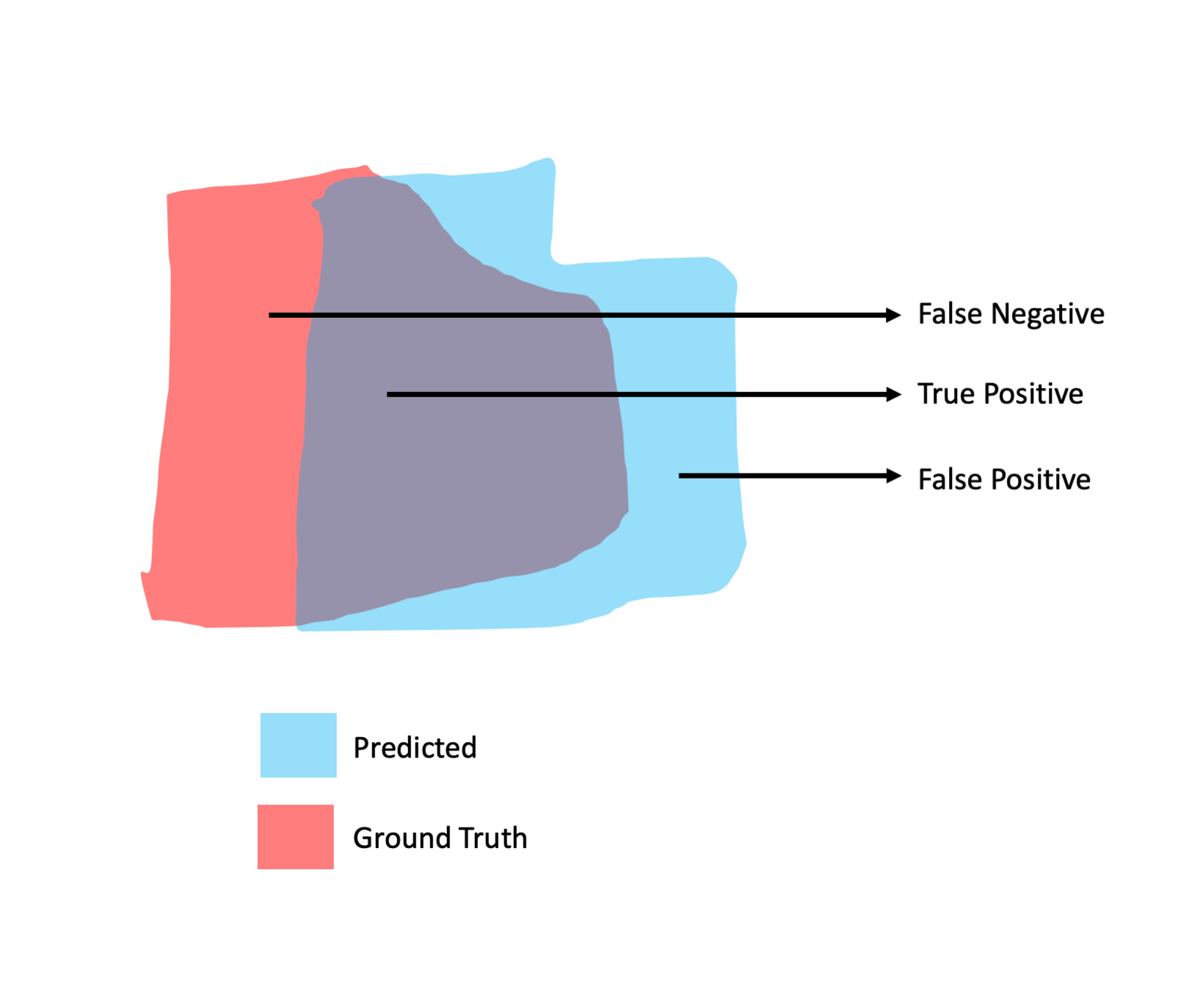}
		\caption{Intersection over Union}
		\label{fig:iou}
	\end{figure}

	\begin{gather}
	\text{IoU} = \frac{\text{TP}}{\text{TP+FN+FP}}
	\end{gather}
	Unlike recall, this metric also penalizes performance based on false positives.
	
	\item{Boundary F1 (BF) contour matching score : }
	In order to check the quality with which the boundary of segmentation is predicted, boundary F1 score is used. Each pixel of the contour of the predicted segmentation map is classified as correct or incorrect based on its distance from the closest point on the ground truth contour. Then recall and precision are calculated. The boundary F1 score is computed as the harmonic mean of recall and precision.
	
	\begin{center}
		BF Score = $\frac{2 \times \text{Precision} \times \text{Recall} }{\text{Recall} + \text{Precision}}$
	\end{center}
	
	BF score lies in the range 0 to 1, where a score of 1 denotes perfect match between the predicted and ground truth segment contours. A threshold known as the distance error tolerance, is used to decide if a predicted segment pixel matches with the ground truth segment. A default value of $0.75\%$ times the diagonal length of the image, i.e., $0.0075 \times 800 \times \sqrt{2} = 8.48$, is used as the threshold.
	
\end{itemize}

\section{Training Settings}
\begin{figure}[H]
	\centering
	\includegraphics[width=0.75\textwidth]{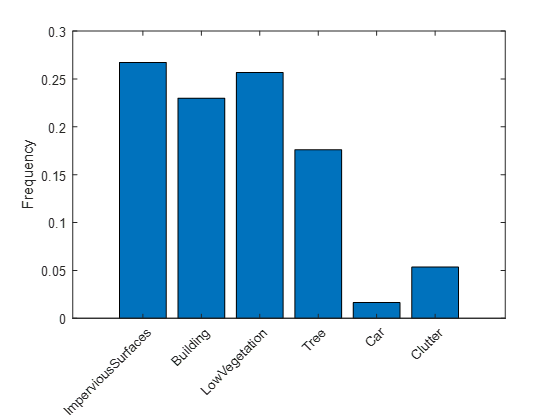}
	\caption{Class weights used during training.}
	\label{fig:class_weights}
\end{figure}

We use DeepLab v3+ \cite{chen2018encoder} for performing semantic segmentation of aerial imagery. For training the network, we used the ISPRS Potsdam dataset \cite{potsdam}. All the labels, namely, \textit{impervious surfaces}, \textit{buildings}, \textit{low vegetation}, \textit{tree}, \textit{car} and \textit{clutter}, were used for training. Each of the tiles in the RGB images and labels of the training set were randomly cropped to patches of size $800 \times 800$ pixels, and then randomly zoomed in and out, flipped and rotated for augmenting the training data. Class weights, as shown in Fig. \ref{fig:class_weights}, were then used to account for class imbalance. These were found by calculating the ratio of the total number of pixels of a class to the total number of pixels in all the training images.

We used \textit{Xception} \cite{chollet2017xception} base network with pre-trained ImageNet \cite{deng2009imagenet} weights. The network was then trained using Adam optimiser for $20$ epochs with a learning rate of $10^{-4}$, mini-batch size of $4$ and $230$ iterations per epoch on a NVIDIA GeForce RTX 2070. We also used $L_2$ regularization with a factor of $0.005$ and shuffled the training data in every epoch. For validation, $25$ \% of the training data was used. 

\section{Test Set Results}
For predicting on a test tile, a sliding window of size $800 \times 800$ pixels with an overlap of $400$ pixels was used to predict the labels and the results were then stitched together. We have shown below the input RGB image, ground truth label and predicted label for two tiles from the test set.

\begin{figure}[H]
	\centering
	\includegraphics[width=0.65\textwidth]{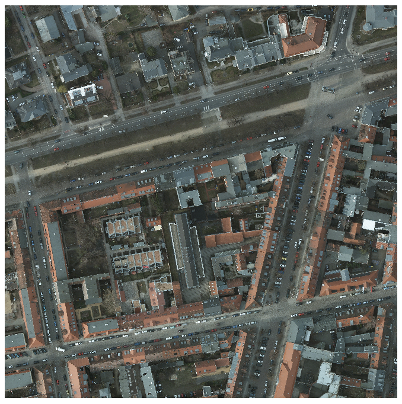}
	\caption{Example 1: RGB Aerial Image.}
	\label{fig:postsdamImg1}
\end{figure}

\begin{figure}[H]
	\centering
	\includegraphics[width=0.8\textwidth]{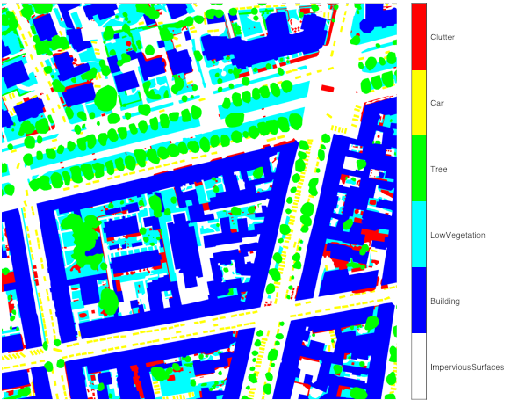}
	\caption{Example 1: Ground Truth Label.}
	\label{fig:postsdamGT1}
\end{figure}

\begin{figure}[H]
	\centering
	\includegraphics[width=0.8\textwidth]{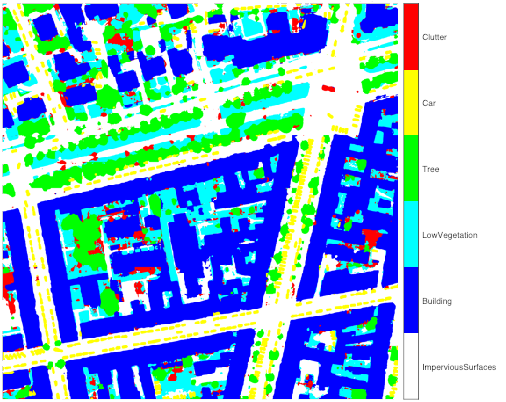}
	\caption{Example 1: Predicted Label.}
	\label{fig:postsdamPred1}
\end{figure}

\begin{figure}[H]
	\centering
	\includegraphics[width=0.65\textwidth]{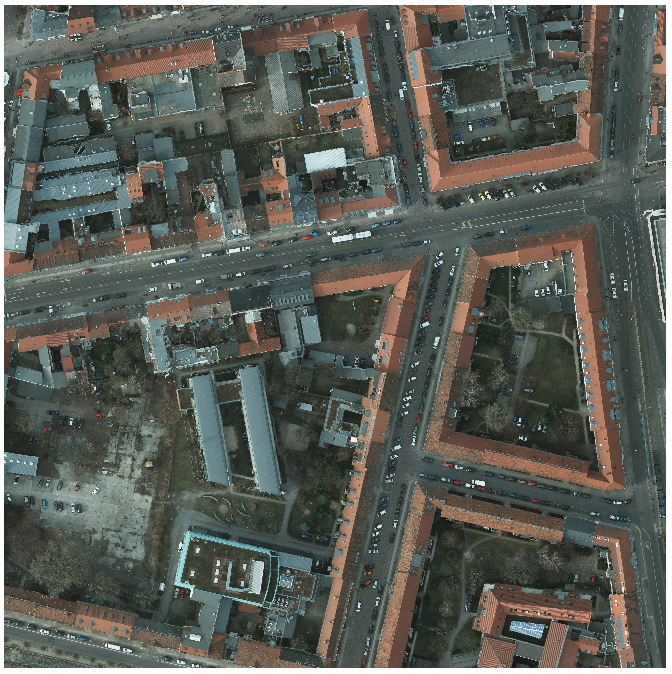}
	\caption{Example 2: RGB Aerial Image.}
	\label{fig:postsdamImg2}
\end{figure}

\begin{figure}[H]
	\centering
	\includegraphics[width=0.8\textwidth]{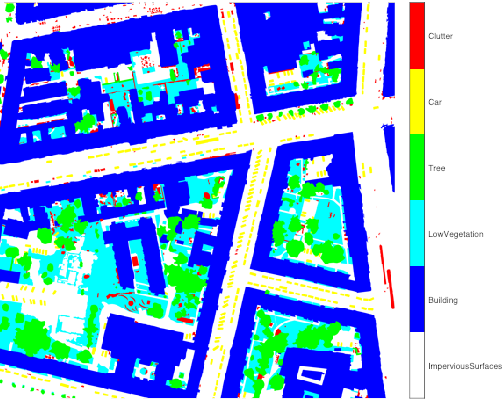}
	\caption{Example 2: Ground Truth Label.}
	\label{fig:postsdamGT2}
\end{figure}

\begin{figure}[H]
	\centering
	\includegraphics[width=0.8\textwidth]{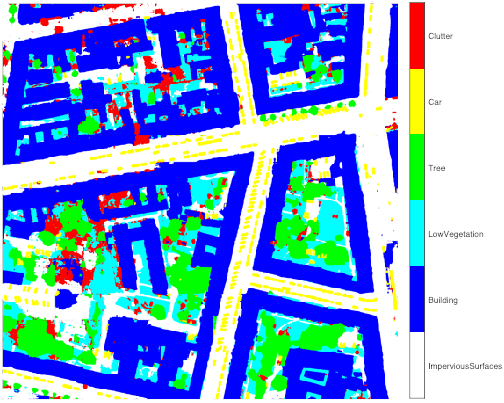}
	\caption{Example 2: Predicted Label.}
	\label{fig:postsdamPred2}
\end{figure}

We report the class wise recall, Intersection over Union (IoU), and Boundary F1 (BF) contour matching score on the test set in Table \ref{tab:tableSegmentation}. 
\begin{table}[H]
	\begin{center}
		\begin{tabular}{|p{4cm}|p{2cm}|p{2cm}|p{2cm}|}
			\hline
			\textbf{Class} & \textbf{Recall} & \textbf{IoU} &  BF \textbf{Score}\\
			\hline
			Impervious Surfaces  & 0.80821 & 0.76650 &  0.92831\\
			\hline
			Building &  0.93768  &   0.87353    &  0.89388 \\
			\hline
			Low Vegetation    & 0.74346  &   0.57438  &    0.80486\\
			\hline
			Tree      &        0.81295  &   0.61912   &   0.89304\\  
			\hline
			Car       &         0.98158   &  0.58406   &   0.94777 \\
			\hline
			Clutter   &        0.59747   &  0.29685    &   0.53180  \\
			\hline 
		\end{tabular}
		\caption{Class wise metrics of Semantic Segmentation.}
		\label{tab:tableSegmentation}
	\end{center}
\end{table}

For the sub-6 GHz carrier frequency case, only a limited set of objects act as blockage. Thus, for the case of LTE, since a carrier frequency of 2 GHz is used in Vienna 5G Simulator, we are mainly concerned with the performance of the model on a subset of classes, namely, \textit{impervious surfaces}, \textit{buildings}, and \textit{tree}. For MMW, it would be required to consider other classes such as vegetation as blockages for calculation of received power by user.

\section{Creating 2.5 D model}
The segmentation map obtained from DeepLabv3+ is used along with DSM to create a 2.5D model of the city. Since each RGB tile in the ISPRS Potsdam dataset \cite{potsdam} is $6000 \times 6000$ pixels, and the pixel scale is $5cm$ per pixel, each tile represents an area of $300m \times 300m$. Vienna 5G System Level Simulator requires blockage dimensions to be in meters, so we scaled the segmentation map by $0.05$. For simulating an area of $900m \times 900m$, we take 9 segmentation maps in a $3 \times 3$ grid. Contour of buildings from each segmentation map is used as the boudary and the corresponding average elevation is used as the height of each building. Each building also has a corresponding ground elevation which is calculated by finding the average DSM elevation in a $50$ pixel, that is, $2.5$  m region around each building segment.

\section{Concluding Remarks}
In this chapter, we introduced semantic segmentation, and covered the most recent concepts and techniques such as atrous spatial pyramid pooling, and encoder-decoder architecture that are used in DeepLabv3+ network. We then presented the evaluation metrics for this task and the results of training the network on the ISPRS Potsdam dataset. We also briefly explained how 2.5D model was created from the segmentation map. In the next chapter, the evaluation metrics for comparing BS placement and results obtained by NSGA-II based optimization have been discussed.

%% file: Chapters/Chapter5.tex
% Chapter Template

\chapter{Evaluation of approach} % Main chapter title
%TP/FP tell
% Precision Recall formula

\label{Chapter5} % Change X to a consecutive number; for referencing this chapter elsewhere, use \ref{ChapterX}

\lhead{Chapter 5. \emph{Evaluation of Approach}} % Change X to a consecutive number; this is for the header on each page - perhaps a shortened title

%----------------------------------------------------------------------------------------
%	SECTION 1
%----------------------------------------------------------------------------------------

The results and analysis of the simulations performed using our approach have been presented in this Chapter. We first introduce metrics for examining base station deployments and then consider two scenarios, one involving placement where no base stations have been deployed so far, and the second involves placement where existing base stations have already been deployed. For each of these scenarios, we provide details of the simulation parameters followed by qualitative and quantitative results to illustrate the performance of our approach. This is followed by effect of using blockages for finding optimal BS location and comparison with other approaches.

%\section{Placement}

\section{Evaluation Metrics}
For evaluating the performance of various placement configurations, we use the same parameter as in \cite{simic2017coverage}, that is, SINR coverage probability. Additionally, we also consider user throughput CDF that can be simulated in Vienna 5G simulator. These metrics are briefly described below.
\begin{itemize}
	\item SINR coverage probability : 
	SINR is defined as the ratio of the power received from the serving BS to the sum of power from non-serving BS and noise power. SINR coverage probability is the probability that a user has SINR above a threshold $P(SINR>\tau)$, where $\tau$ is the threshold. It is the Complementary Cumulative Distribution Function (CCDF) of SINR. A higher SINR coverage probability implies better coverage.
	\item CDF of user downlink throughput : 
	Throughput is defined as the number of information bits successfully received by a user per unit time. Its CDF represents the probability that a user has a throughput lower than a threshold $P(throughput<t)$. A lower user downlink throughput CDF implies better data speed.
\end{itemize}

We consider two regions, Scenario I and II, each of area $900m \times 900m$ from the ISPRS Potsdam dataset. The RGB tiles were segmented using DeepLabv3+ to generate blockages and then NSGA-II was used to find the optimal base station locations. Initial population size was set as 48 and the optimizer was run for 40 iterations. The crossover probability was taken as 0.9 and the mutation probability was taken 0.99. 

\section{Placement without Prior Deployed BS}
\label{sec:placenoprior}
 First, we considered the case when no existing base stations were present. The maximum number of base stations to be deployed, $M_{max}$, was set to 6. 
\begin{itemize}
	\item Scenario I : \\
	In the first scenario, we considered the tiles labelled as $5\_11$, $5\_12$, $5\_13$, $6\_11$, $6\_12$, $6\_13$, $7\_11$, $7\_12$, $7\_13$ in the Potsdam dataset. The top view of the 2.5D model with candidate BS locations (red), and users (yellow: users near buildings and on roads, green: other users) generated have been shown in Fig. \ref{fig:nsga2opta}. 
	\begin{figure}[H]%
		\includegraphics[height=4in]{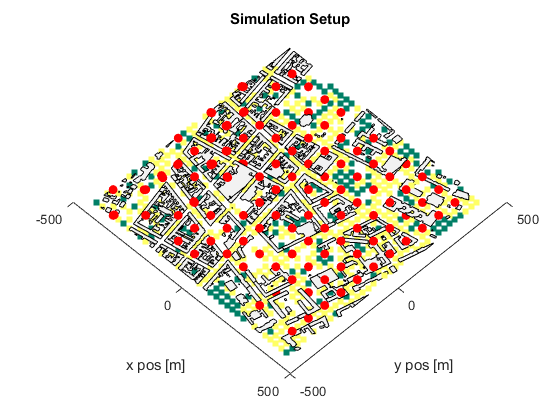}%
		\caption{Top view of Scenario I with candidate BS location}%
		\label{fig:nsga2opta}
	\end{figure}

	\begin{figure}[H]%
		\includegraphics[height=3.5in]{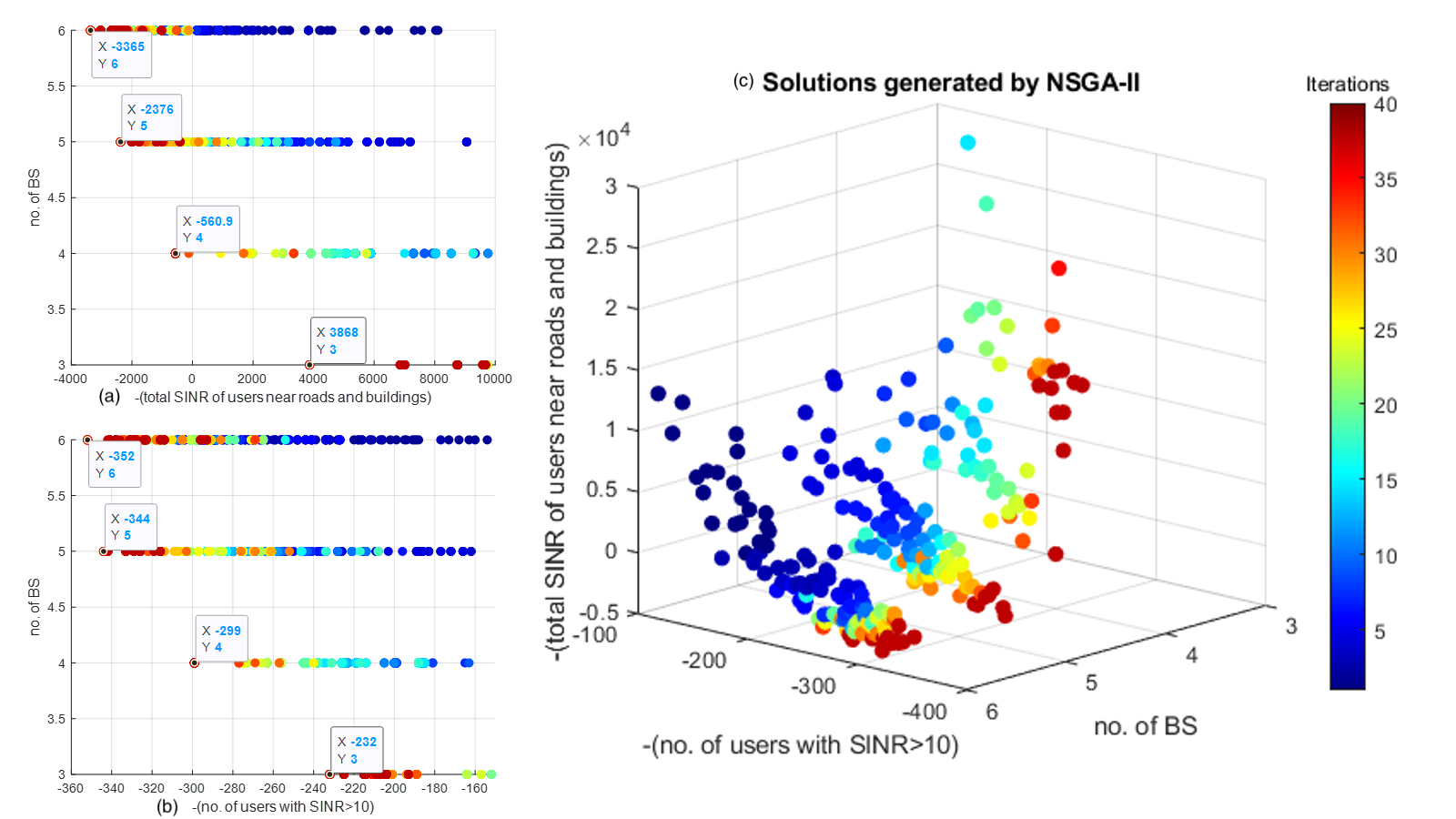}%
		\caption{Scenario I Solutions generated by NSGA-II optimizer}%
		\label{fig:nsga2optb}
	\end{figure}

	The results of NSGA-II optimizer have been shown in Fig.  \ref{fig:nsga2optb}. The plot in (c) shows the solutions generated by the optimizer with the color varying with every iteration. The axes represent the objective functions used in our formulation in (\ref{sec:objfun}). It can be seen that it generates a better solution with every iteration. The 2D versions of this plot have been shown on the left for comparison. The total SINR, i.e., sum of SINR of the users near buildings and on roads increases with the number of base stations deployed as seen in (a). Similarly in (b), the number of users with $SINR > 10$ dB also follows the same trend. 
	
	\begin{figure}[H]%
	\includegraphics[height=4.5in]{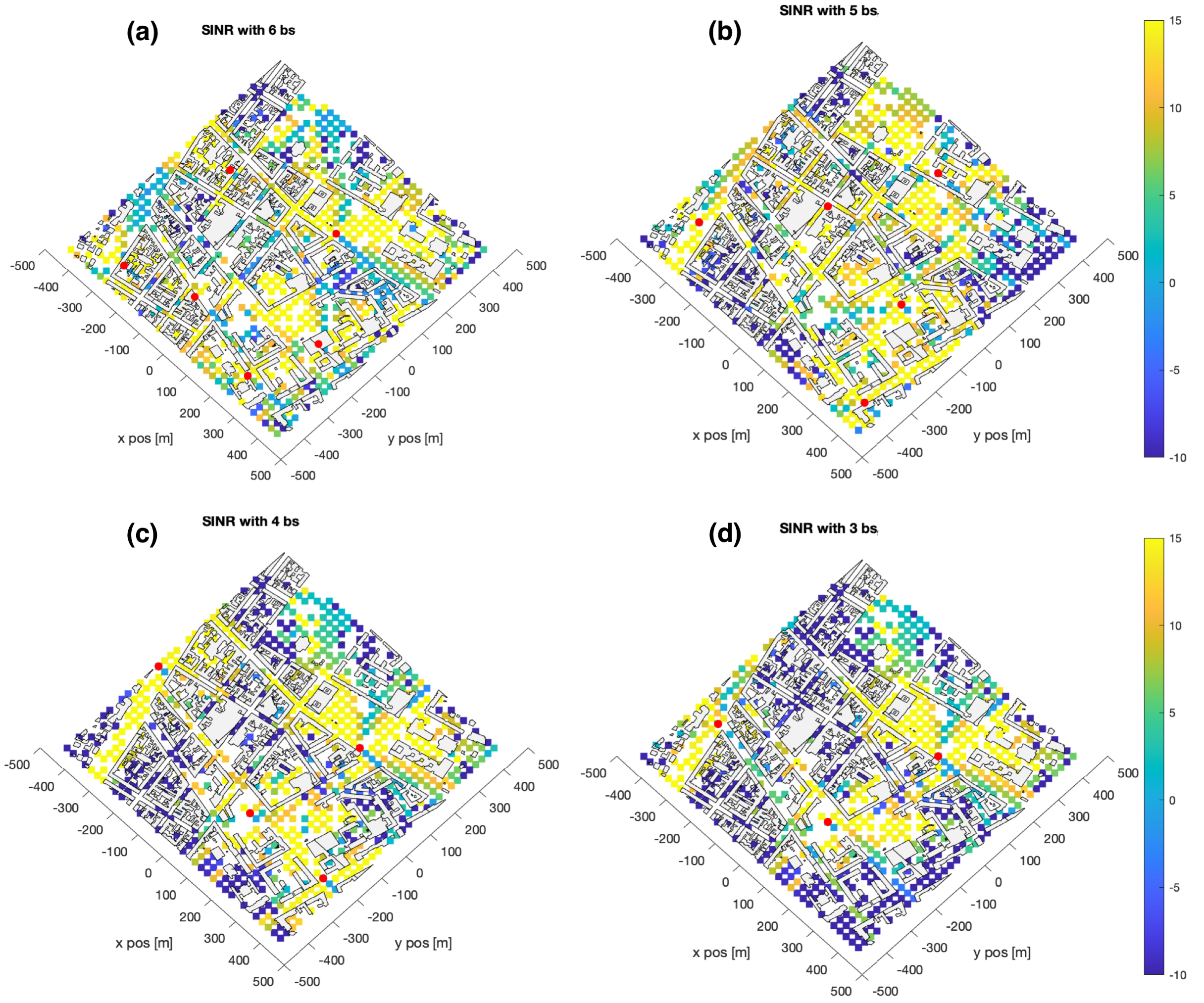}%
	\caption{Scenario I Solution Heatmap}%
	\label{fig:nsga2optc}
	\end{figure}

	In Fig. \ref{fig:nsga2optc}, the optimal solutions for number of base stations = 3, 4, 5 and 6 from the final population based on non-dominated sorting were chosen and plotted. 
%	It can be seen that, for the configurations shown in Fig. \ref{fig:nsga2optc}, both SINR coverage probability and user downlink throughput improved by increasing the number of base stations. 
%		
	\begin{figure}[H]
		\centering
		\includegraphics[width=0.6\textwidth]{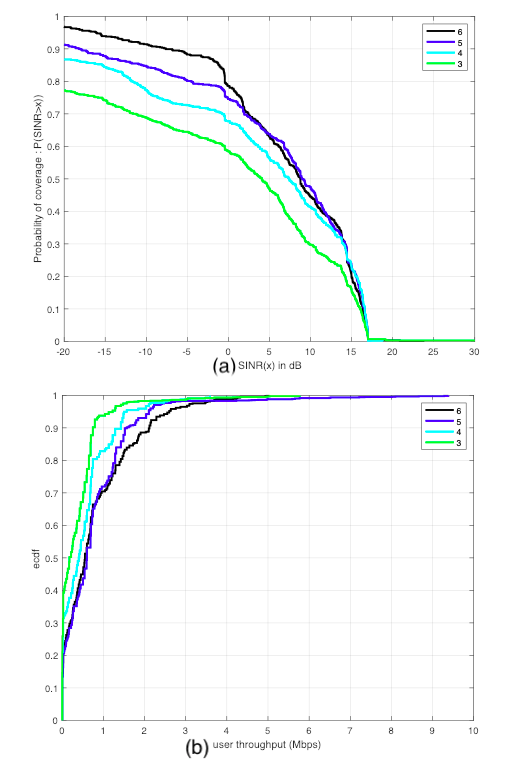}
		\caption{Comparison of placement configurations for Scenario I.}
		\label{fig:nsga2optlink}
	\end{figure}
	
	We then performed link level simulations for each of these configurations and plotted the SINR coverage probability and user downlink throughout CDF using Vienna 5G System level Simulator as shown in Figs. \ref{fig:nsga2optlink} (a) and (b) respectively. In (a), we observe that the SINR coverage probability increases as number of BS increases from 3 to 5. For 6 BS, however, the probability of coverage above 0 dB is almost identical to the case of 5 BS. Therefore, we infer that deploying 6 BS offers almost identical coverage to deploying 5 BS. Further, in (b), we see that the user downlink throughput CDF decreases with increasing number of BS. This implies that deploying more BS decreases the probability of a user having a lower throughput. We again observe a high degree of overlap for 5 and 6 BS cases. Therefore, for the given scenario, we conclude that deploying 5 or 6 BS would result in similar coverage, however, 6 BS will have a higher throughput for users.

	\item Scenario II :\\ 
	In the second scenario, we consider the $900m \times 900m$ area comprising of tiles labelled as $2\_10, 2\_11, 2\_12, 3\_10, 3\_11, 3\_12, 4\_10, 4\_11, 4\_12$ in the Potsdam dataset. The top view of the 2.5D model with candidate BS locations (red), and users (yellow: users near buildings and on roads, green: other users) generated have been shown in Fig. \ref{fig:nsga2opt2a}. 
	\begin{figure}[H]%
		\centering
		\includegraphics[height=4in]{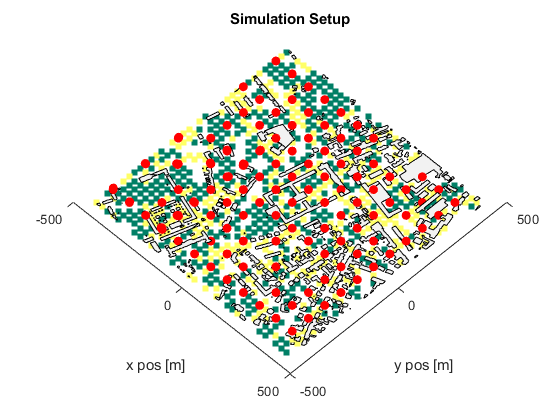}%
		\caption{Top view of Scenario I with candidate BS location}%
		\label{fig:nsga2opt2a}
	\end{figure}

	The results of NSGA-II optimizer have been shown in Fig.  \ref{fig:nsga2opt2b}. The plot in (c) shows the solutions generated by the optimizer with the color varying with every iteration. It can be seen that it generates an optimal solution with every iteration. The total SINR of the users near buildings and on roads increases with the number of base stations deployed as seen in (a). Similarly in (b), the number of users with $SINR > 10$ dB also follows the same trend. 

	\begin{figure}[H]%
		\centering
		\includegraphics[height=3.5in]{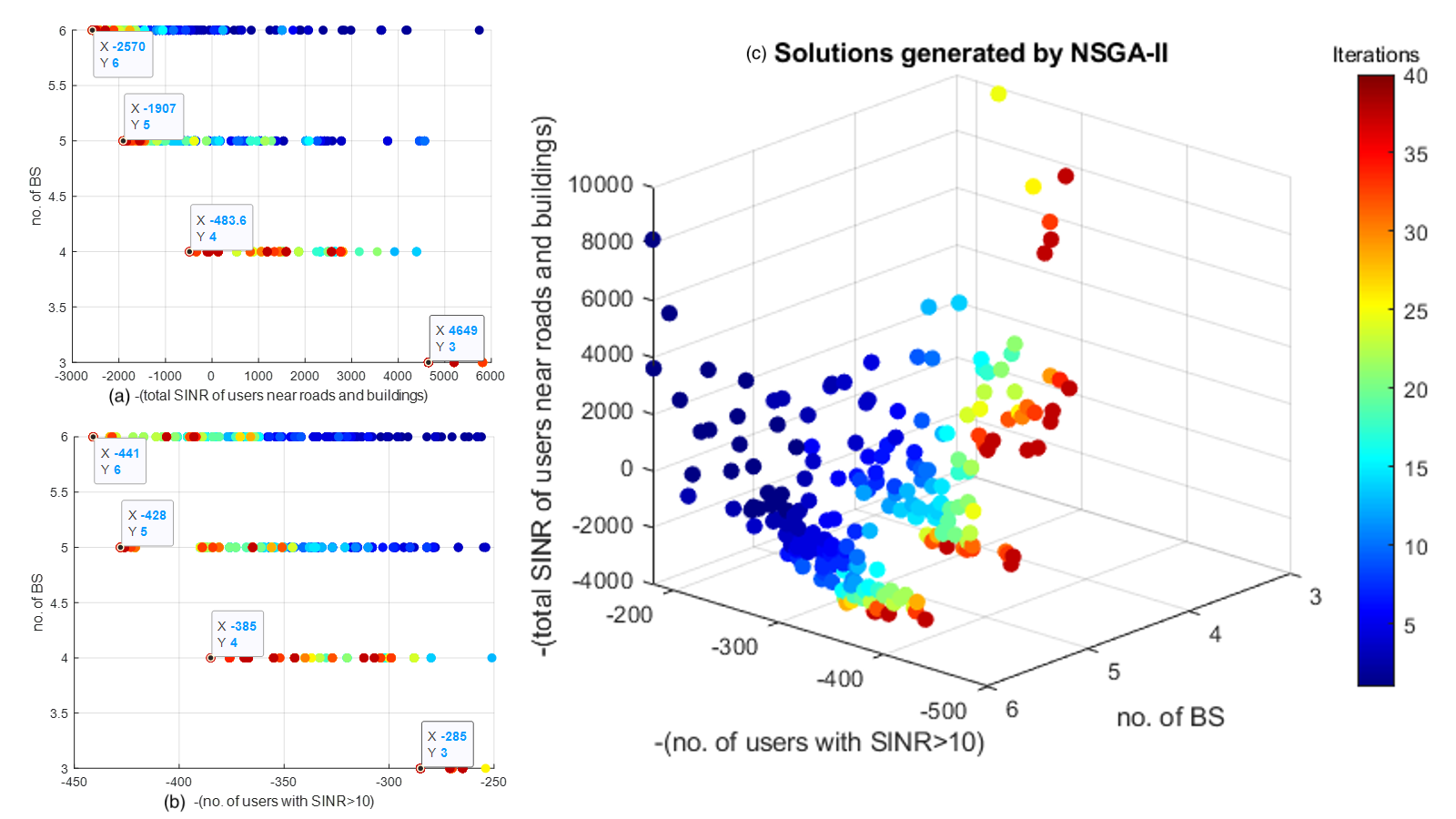}%	
		\caption{Scenario II Solutions generated by NSGA-II optimizer}%
		\label{fig:nsga2opt2b}
	\end{figure}
	
	\begin{figure}[H]%
		\centering
		\includegraphics[height=4.5in]{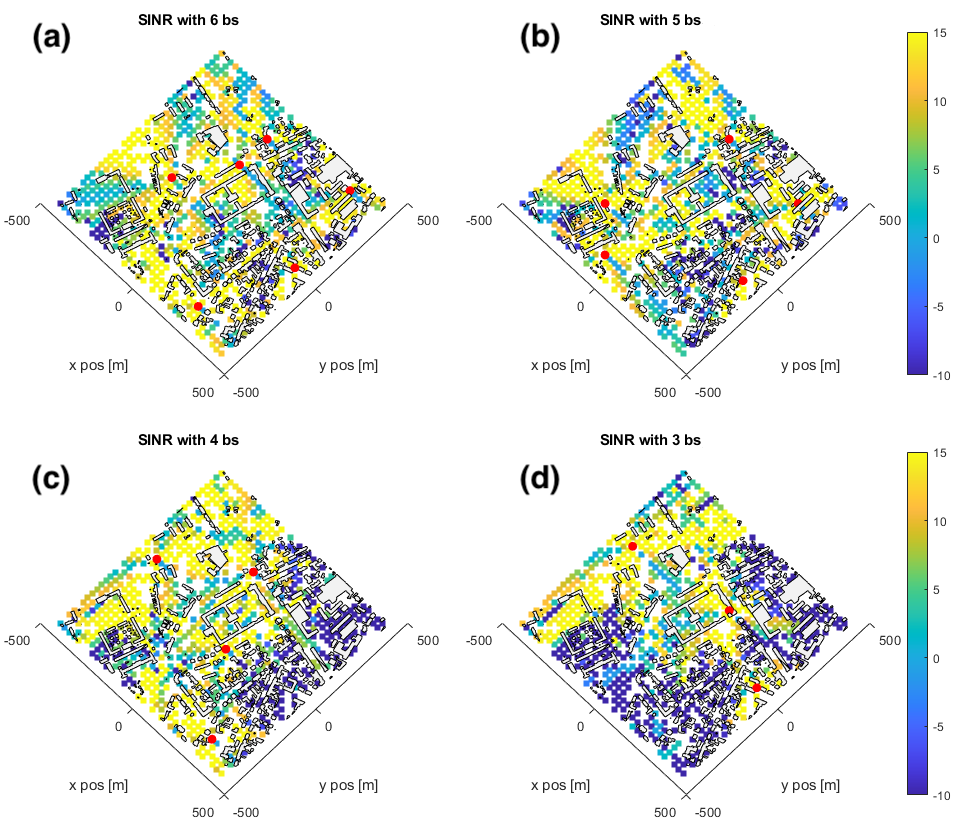}%
		\caption{Scenario II Solution Heatmap}%
		\label{fig:nsga2opt2c}
	\end{figure}

	In Fig. \ref{fig:nsga2opt2c}, the optimal solutions for number of base stations = 3, 4, 5 and 6 from the final population based on non-dominated sorting were chosen and plotted. We then performed link level simulations for each of these configurations and plotted the SINR coverage probability and user downlink throughput CDF using Vienna 5G System level Simulator as shown in Figs. \ref{fig:nsga2opt2link} (a) and (b) respectively. For the configurations shown in Fig. \ref{fig:nsga2opt2c}, both coverage probability and user downlink throughput improved by increasing the number of base stations as seen in Scenario I.  Similar conclusions can be drawn regarding the optimal number of base stations to be deployed as in Scenario I, the difference being that in Scenario II, the SINR coverage probability for 5 and 6 BS cases do not overlap significantly when compared to Scenario I. 
	
	\begin{figure}[H]
		\centering
		\includegraphics[width=0.5\textwidth]{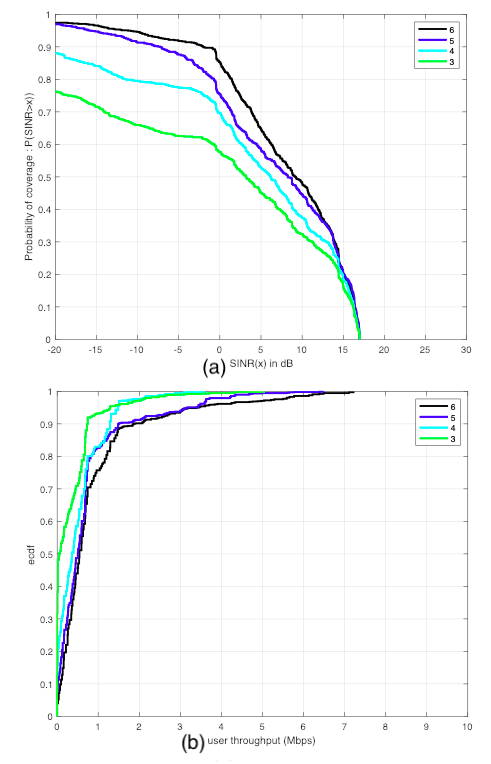}
		\caption{Comparison of placement configurations for Scenario II.}
		\label{fig:nsga2opt2link}
	\end{figure}
	
\end{itemize}

\section{Placement with Prior Deployed BS}
\begin{figure}[H]%
	\centering
	\includegraphics[height=4in]{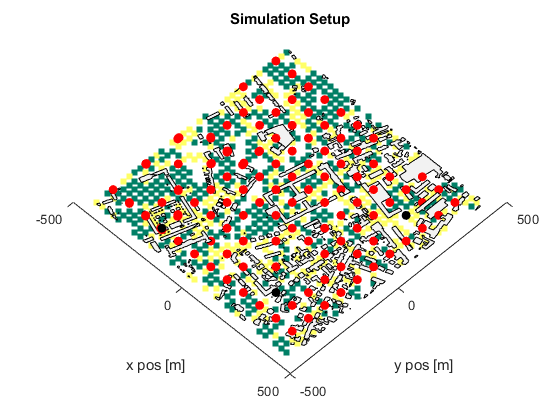}%
	\caption{Top view of Scenario II with candidate and existing BS location}%
	\label{fig:nsga2opt3a}
\end{figure}

We also consider the case when new base stations are to be deployed given the location of existing base stations. Here, NSGA-II only finds optimal locations for the new base stations. We consider the same scenario as Scenario II as described in section \ref{sec:placenoprior}, but with existing base stations. 
We assumed that 3 base stations (represented by black) were already present as shown in Fig. \ref{fig:nsga2opt3a}. As earlier, red represents candidate base stations, yellow and green represents users nears buildings and on roads and other users respectively. 
 
\begin{figure}[H]%
	\centering
	\includegraphics[width=1\textwidth]{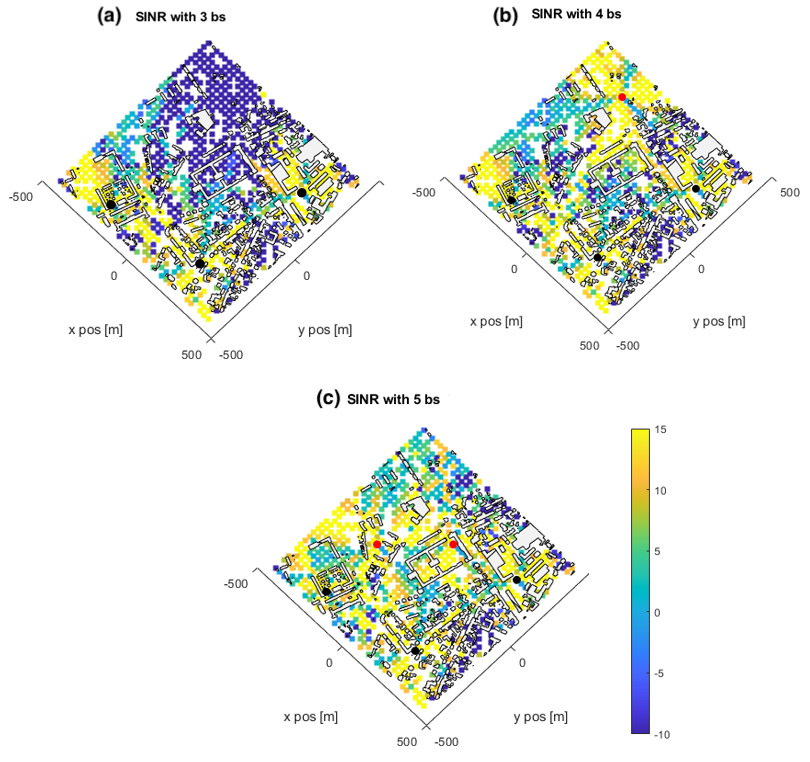}%
	\caption{Scenario I Solutions generated by NSGA-II with prior BS}%
	\label{fig:nsga2opt3b}
\end{figure}

After using NSGA-II and convergence of the optimization, the optimal solution for total number of base stations = 3, 4, and 5 from the final population based on non-dominated sorting were chosen and plotted in Fig. \ref{fig:nsga2opt3b}. 

\begin{figure}[H]
	\centering
	\includegraphics[width=0.65\textwidth]{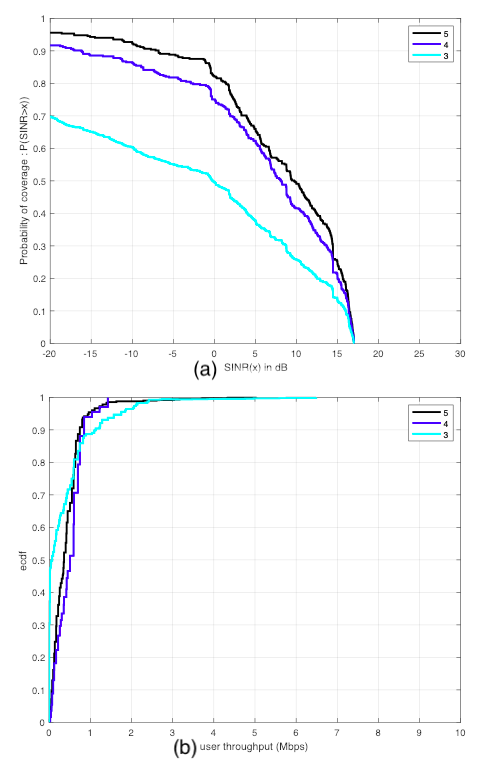}
	\caption{Effect of deploying new BS.}
	\label{fig:nsga2opt3c}
\end{figure}

We then performed link level simulations for each of these configurations and plotted the SINR coverage probability and user downlink throughput CDF using Vienna 5G System level Simulator as shown in Figs. \ref{fig:nsga2opt3c} (a) and (b) respectively. The SINR coverage probability improves by deploying more new base stations as shown in Fig. \ref{fig:nsga2opt3c} (a). \\

In Fig. \ref{fig:nsga2opt3c} (b) part, we see that for 3 BS, the probability of a user having no throughput ($P(throughput<0Mbps)$) is 0.5. This is supported by the plot in Fig. \ref{fig:nsga2opt3b} (a), where we can see that about half of the users have no coverage (very low SINR). For 4 BS and 5 BS, $P(throughput<0Mbps)$ decreased to almost 0. However, the trend of a lower throughput CDF for more number of BS is not seen uniformly in this example. Unlike the previous section (\ref{sec:placenoprior}), this is potentially because of the presence of existing BS which are not at their optimal positions.\\

Therefore for Scenario II, placing all BS using NSGAII resulted in improved SINR coverage probability and user downlink throughput. But when 3 BS were already present and remaining BS are placed using NSGAII, only SINR coverage probability improved.

\section{Effect of Using Blockages for Optimization}
\begin{figure}[H]
	\centering
	\includegraphics[width=1\textwidth]{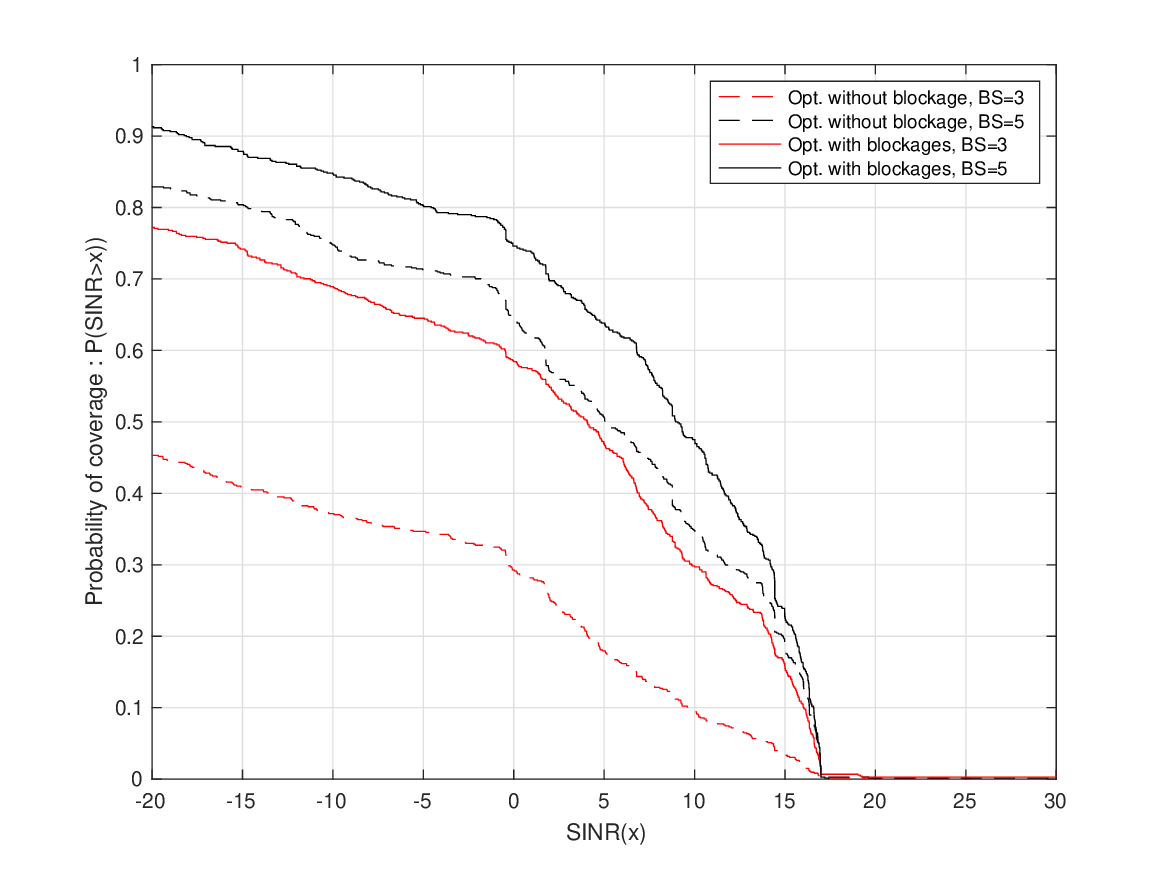}
	\caption{Effect of considering blockages in optimal BS placement.}
	\label{fig:nsgaopt_no_semantic}
\end{figure}

To quantify the effect of considering blockages such as buildings for calculation of SINR during optimization, we consider the same  example (Scenario I) in sub-section \ref{sec:placenoprior}. However, this time we once perform the optimization, once with blockages (NSGAII (1)) and without considering any blockages (NSGAII (2)). The ground surface elevation was used for simulating the scenario in both the cases. After obtaining the optimal base station locations, we plotted the SINR coverage probability for 3 and 5 base stations as shown in Fig. \ref{fig:nsgaopt_no_semantic}.

\begin{figure}[H]
	\centering
	\includegraphics[width=1\textwidth]{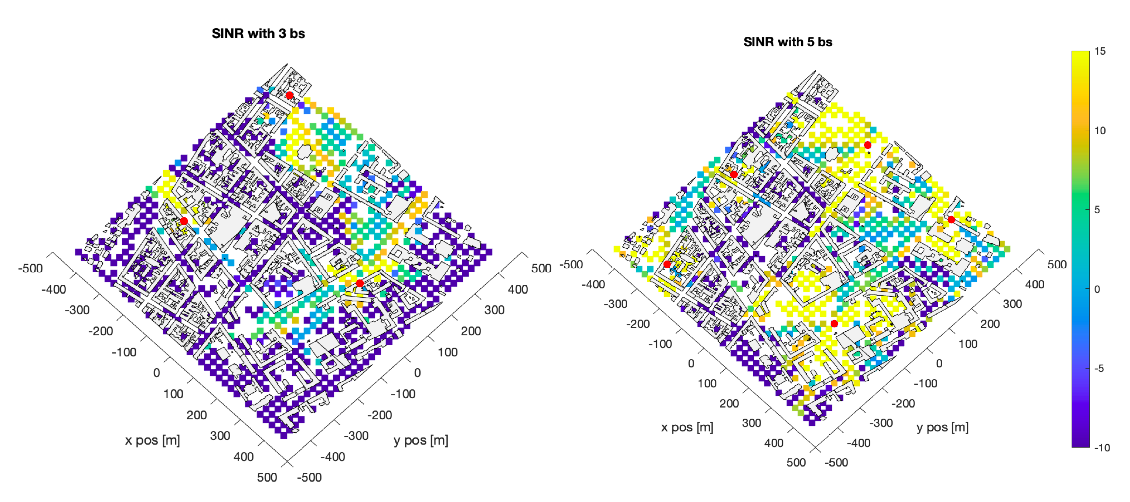}
	\caption{Optimal BS location when blockages are not considered during optimization.}
	\label{fig:nosemantic_bs}
\end{figure}

It can be seen that by considering blockages for finding optimal base station location, the SINR coverage probability improved as compared to not considering them during optimization. The corresponding base station locations (for the dashed red and black lines) have been shown in Fig. \ref{fig:nosemantic_bs}. As blockages were not considered, the optimal base station locations form a convex polygon unlike in Fig. \ref{fig:nsga2optc} (b) and (d).

\section{Comparison with Other Methods}
We compare our NSGA-II based placement approach to the iterative k-means approach proposed by Toros \textit{et.al.} in \cite{torHos2010algorithm} as discussed in subsection \ref{sec:litusr}. The SINR coverage probability for optimal placement of 3,4 and 5 base stations from NSGA-II based optimization and iterative k-means has been shown in Fig. \ref{fig:kmeans_vs_nsga2}. We have considered the same Scenario I as in subsection \ref{sec:placenoprior} for this comparison.

\begin{itemize}
	\item Iterative k-means approach:
	
	Initially a single base station was placed at the center of the scenario and the Most Unserved Sector (MUS) was identified as the sector with the minimum number of users with $SINR < 10 dB$ in that sector. A new base station was then placed in the MUS and k-means was used to cluster all the user locations with $k=2$ (number of base stations) and initial cluster centers as the original and the newly added base station location. The newly obtained cluster centroids were the optimal location of placing 2 base stations. Then, MUS was found again and this process was continued till 5 base stations were placed optimally.
	
	\begin{figure}[H]
		\centering
		\includegraphics[width=1\textwidth]{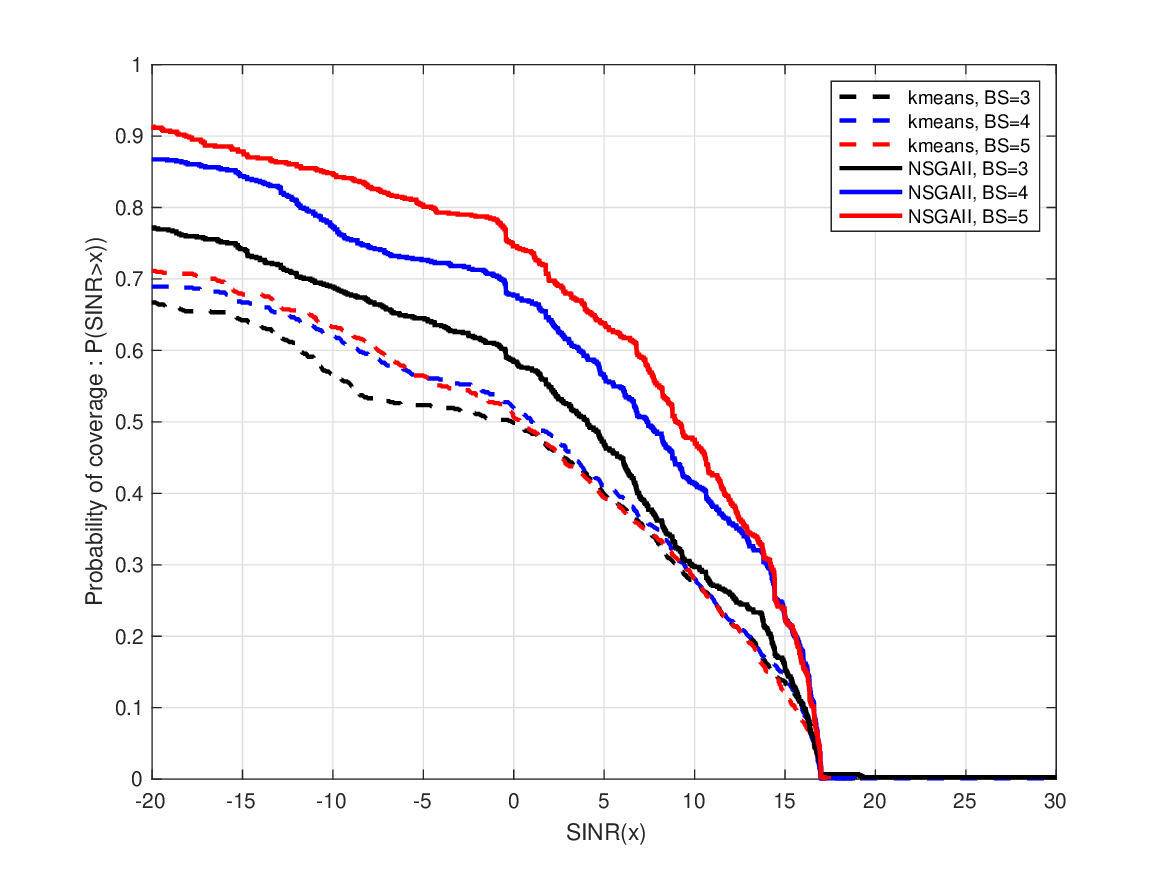}
		\caption{Comparision with iterative k-means method.}
		\label{fig:kmeans_vs_nsga2}
	\end{figure}
	
	From Fig. \ref{fig:kmeans_vs_nsga2}, we can see that despite increasing the number of base stations, the SINR coverage probability did not improve significantly for placement using the iterative k-means approach. This can be attributed to the lack of using SINR while computing the optimal base station location. The method uses SINR only for finding MUS.\\ 
	
	We also observe that for the scenario under consideration, 3 base stations placed using NSGA-II provide better SINR coverage probability than 5 base stations placed using k-means approach.

	\item Standard Genetic Algorithm: 
	
	We only maximized the number of users with $SINR > 10 dB$ as mentioned in (\ref{eq:obj2}) since standard GA is single objective. The rest of the parameters were kept the same as used in NSGA-II. For finding the optimal location of deploying 3,4 and 5 base stations, we separately ran the GA optimizer thrice by varying $M_{max}$ each time. The SINR coverage probability comparing optimal placement of 3,4 and 5 base stations using GA and NSGA-II have been shown in Fig. \ref{fig:ga_vs_nsga2}.
	
	\begin{figure}[H]
		\centering
		\includegraphics[width=1\textwidth]{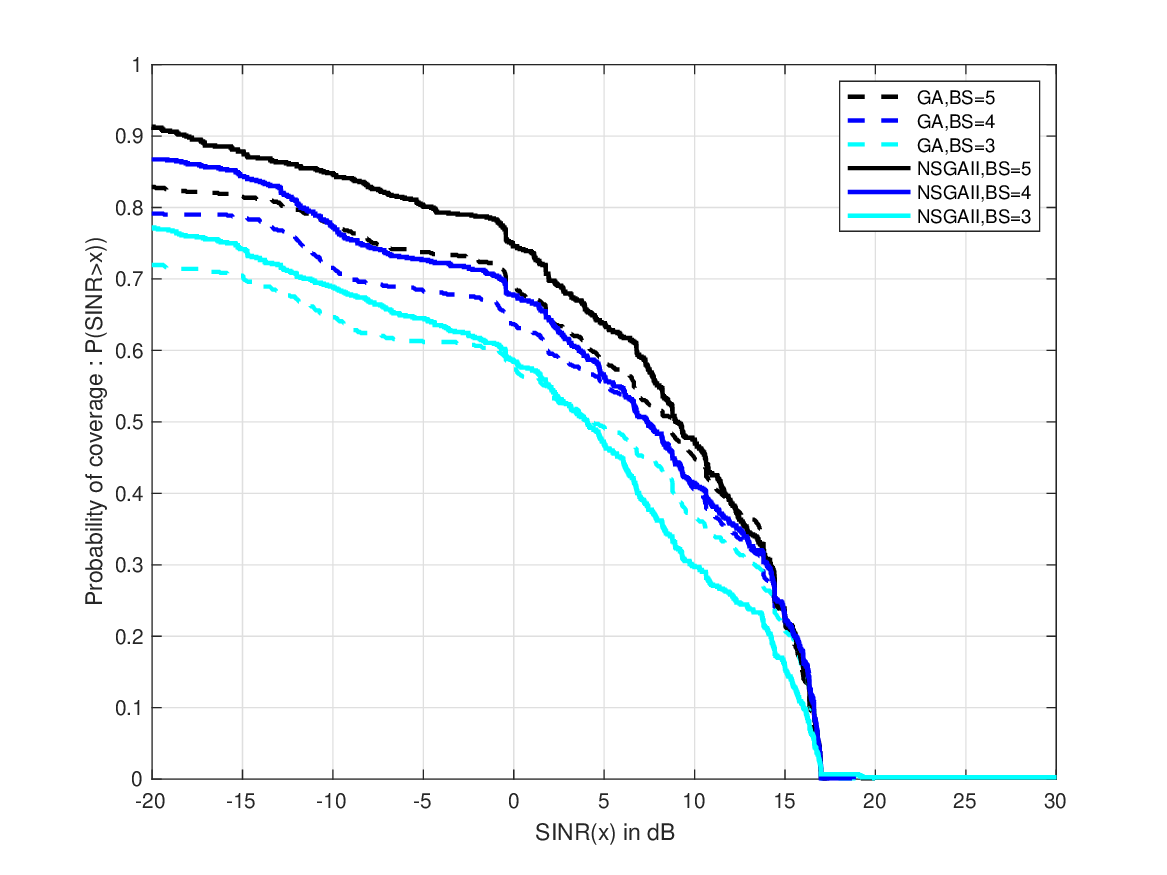}
		\caption{Comparision with normal GA.}
		\label{fig:ga_vs_nsga2}
	\end{figure}

	From the figure, we can see that placement using NSGA-II gives slightly higher SINR coverage probability as compared to GA for 5 and 4 base stations. This improvement is not seen for 3 base stations. Therefore, the benefit of multi-objective optimization is only considerable when optimal number of base stations are deployed.\\
	
	Another observation is that, single objective optimization with standard GA gives higher SINR coverage probability as compared to the iterative k-means approach shown in Fig. \ref{fig:kmeans_vs_nsga2}. 

\end{itemize} 

We also compare all the placement methods discussed so far in terms of the total SINR of all users and number of users with SINR greater than a threshold, that is, objective functions mentioned in ( \ref{eq:obj1}) and (\ref{eq:obj3}). This has been shown in Table \ref{tab:table2}. NSGAII $(1)$ refers to optimization with blockages and NSGAII $(2)$ refers to optimization without blockages taken into consideration. 
 
\begin{table}[hbtp]
	\begin{center}
			\begin{tabular}{|p{4cm}|p{2cm}|p{2cm}|p{2cm}|p{2cm}|}
				\hline
				\multicolumn{1}{|c|}{Method} & \multicolumn{1}{|c|}{K-means \cite{torHos2010algorithm} }  & 
				\multicolumn{1}{|c|}{GA} & 
				\multicolumn{1}{|c|}{NSGAII (2)}& 
				\multicolumn{1}{|c|}{\textbf{NSGAII (1)}}\\%without blockage
				\hline
				\multicolumn{5}{|c|}{BS=3}\\
				\hline
				$\sum_{k} SINR_k$  & $-1.580 X 10^4$ & $-1.616 X 10^4$ & $-2.937 X 10^4$ & $-3.868 X 10^3$ \\
				\hline
				$\sum_{i=1}^{N} \mathbbm{I}{ (SINR_i>10dB)}$  & 200 & 265 & 69 & 232\\
				\hline
				
				\multicolumn{5}{|c|}{BS=4}\\
				\hline
				$\sum_{k} SINR_k$  & $-1.134 X 10^4$ & $-4.792 X 10^3$ & $-2.213 X 10^4$ & $5.609 X 10^2$\\
				\hline
				$\sum_{i=1}^{N} \mathbbm{I}{ (SINR_i>10dB)}$  & 203 & 294 & 216 & 299\\
				\hline
				
				\multicolumn{5}{|c|}{BS=5}\\
				\hline
				$\sum_{k} SINR_k$  & $-9.403 X 10^3$ & $-2.216 X 10^3$ & $-2.354 X 10^3$& $2.376 X 10^3$\\
				\hline
				$\sum_{i=1}^{N} \mathbbm{I}{ (SINR_i>10dB)}$  & 204 & 325 & 252 & 344\\
				\hline
				
			\end{tabular}
		
		\caption{Comparison with other methods}
		\label{tab:table2}
	\end{center}
\end{table}
% 
%\begin{table}[hbtp]
%	\begin{center}
%		\begin{tabular}{|p{4cm}|p{2cm}|p{2cm}|p{2cm}|p{2cm}|}
%			\hline
%			\multicolumn{1}{|c|}{Method} & \multicolumn{1}{|c|}{K-means \cite{torHos2010algorithm} }  & 
%			\multicolumn{1}{|c|}{GA} & 
%			\multicolumn{1}{|c|}{NSGAII (2)}& 
%			\multicolumn{1}{|c|}{\textbf{NSGAII (1)}}\\%without blockage
%			\hline
%			\multicolumn{5}{|c|}{BS=3}\\
%			\hline
%			$\sum_{k} SINR_k$  & $1.580 X 10^4$ & $1.616 X 10^4$ & $2.937 X 10^4$ & $3.868 X 10^3$ \\
%			\hline
%			$\sum_{i=1}^{N} \mathbbm{I}{ (SINR_i>10dB)}$  & 200 & 265 & 69 & 232\\
%			\hline
%			
%			\multicolumn{5}{|c|}{BS=4}\\
%			\hline
%			$\sum_{k} SINR_k$  & $1.134 X 10^4$ & $4.792 X 10^3$ & $2.213 X 10^4$ & $5.609 X 10^2$\\
%			\hline
%			$\sum_{i=1}^{N} \mathbbm{I}{ (SINR_i>10dB)}$  & 203 & 294 & 216 & 299\\
%			\hline
%			
%			\multicolumn{5}{|c|}{BS=5}\\
%			\hline
%			$\sum_{k} SINR_k$  & $9.403 X 10^3$ & $2.216 X 10^3$ & $2.354 X 10^3$& $2.376 X 10^3$\\
%			\hline
%			$\sum_{i=1}^{N} \mathbbm{I}{ (SINR_i>10dB)}$  & 204 & 325 & 252 & 344\\
%			\hline
%			
%		\end{tabular}
%		\caption{Comparison with other methods}
%		\label{tab:table2}
%	\end{center}
%\end{table}

Overall, we see that the iterative k-means method does not improve SINR coverage with increasing number of base stations. Standard single objective GA performs better than the iterative k-means method. Multi-objective NSGAII $(1)$ based placement by considering blockages works the best, especially when optimal number of base stations are deployed. The benefit of using NSGAII for optimal placement becomes useful only when attenuation in signal strength due to blockages is considered.

\section{Concluding Remarks}
In both the scenarios that we considered, we investigated the BS deployment cost versus coverage tradeoff. The optimal location for installing new base stations given the location of existing base stations was found. Improvement in SINR coverage probability by considering blockages for optimal base station placement was also illustrated. Our NSGA-II based placement method was also compared and shown to work better than an iterative k-means approach and standard GA. In the next chapter, we discuss conclusions from the proposed pipeline, its limitations and future work directions.

%% file: Chapters/Chapter6.tex
% Chapter Template

\chapter{Conclusions and Future Work} % Main chapter title

\label{Chapter6} % Change X to a consecutive number; for referencing this chapter elsewhere, use \ref{ChapterX}

\lhead{Chapter 6. \emph{Conclusions and Future Work}} % Change X to a consecutive number; this is for the header on each page - perhaps a shortened title

%----------------------------------------------------------------------------------------
%	SECTION 1
%----------------------------------------------------------------------------------------
In this thesis, we proposed a pipeline for optimally placing mobile base stations by considering semantic information extracted from aerial drone imagery using deep learning. Generally, optimal deployment of base stations is not feasible due to geographical limitations. However, our approach takes this into account by performing semantic segmentation of aerial drone imagery.

We introduced the problem statement in Chapter \ref{Chapter1} and in Chapter \ref{Chapter2} provided a literature survey of various approaches used by researchers and industry practioners for optimal placement of base stations to improve wireless network coverage. Next, we explained the two main parts of our proposed pipeline, namely, NSGA-II and DeepLabv3+ in Chapter \ref{Chapter3} and Chapter \ref{Chapter4} respectively. We trained DeepLabv3+ using the ISPRS Potsdam dataset and solved the placement problem as a multi-objective optimization problem. The results and analysis of this were presented in Chapter \ref{Chapter5} where we found optimal base station placement locations for two scenarios in Potsdam city for an LTE network. 

The main contribution of this thesis was in formulating BS placement as a multi-objective problem. This is useful when it is desired to optimally locate the BS, to meet the multiple constraints of specified SINR coverage and user throughput probability, and of quantifying the improvement in SINR coverage and user throughput probability when the BS are placed optimally. The problem of optimally placing additional BS in an environment containing some optimally or otherwise placed pre-existing BS was also addressed.

Secondly, we considered the effect of blockages for finding optimal BS deployment location in LTE network which is useful when the density of buildings is high.

\section{Limitations and Future Work\label{sec:conc}}
Further improvements are required in both parts of our proposed pipeline. Our current segmentation model might be biased, for instance, it might not be able to segment buildings with metallic roofs since it has been trained only on the ISPRS Potsdam dataset. If a road is covered by a canopy, then such a road might not be segmented properly, because of which the BS placement suggested by algorithm would be sub-optimal. Aerial imagery collected from different locations should be used and ablation studies of our semantic segmentation model needs to be performed.

We are using only aerial RGB orthophotos from which 3D model of a scenario cannot be simulated. Using a 3D model for checking line of sight would be more preferrable for simulating higher frequency channels such as MMW. In future, 3D aerial data could be collected and 3D semantic segmentation could be performed.

We have only considered the optimal BS placement problem for LTE. Many telecom operators will deploy BS using MMW as carrier frequency in 5G. MMW is also affected by environmental factors such as weather and our current approach does not account for such variations. Additionally, the effect of beam-steering would also need to be modelled. This could be done by using NYUSUM which supports MMW channel models. However, it would be required to modify the simulator to perform coverage anaylsis for a given scenario. 

Actual attenuation of signal depends on the material of the wall. Our current method assumes the same material loss for all buildings. One approach to solve this would be to train a semantic segmentation model with classes belonging to different material loss, such as glass, concrete buildings etc.

The effect of static blockages such as buildings have only been considered in this work. MMW would also be affected by dynamic blockages such as vehicles, moving objects etc. In future, this should also be taken into account in the optimization problem.

%% file: Bibliography.bib
@book{parsons2000mobile,
	title={The mobile radio propagation channel},
	author={Parsons, John David},
	year={2000},
	publisher={John Wiley {\&} Sons Ltd},
	address="New York"
}

@misc{zetterberg20052003,
	title={{IST-2003-507581 WINNER D5. 4 v. 1.4 Final Report on Link Level and System Level Channel Models}},
	author={Zetterberg, Mats Bengtsson and Yu, Kai and Jald{\'e}n, Niklas and Rautiainen, Terhi and Kalliola, Kimmo and Milojevic, Marko and Schneider, Christian and Hansen, Jan},
	year={2005},
	publisher={Nov}
}

@article{meinila2009winner,
	title={{WINNER II channel models}},
	author={Meinil{\"a}, Juha and Ky{\"o}sti, Pekka and J{\"a}ms{\"a}, Tommi and Hentil{\"a}, Lassi},
	journal={Radio Technologies and Concepts for IMT-Advanced},
	pages={39--92},
	year={2009},
	publisher={Wiley Online Library}
}

@inproceedings{li2016real,
	title={{Real-time UAV weed scout for selective weed control by adaptive robust control and machine learning algorithm}},
	author={Li, Liujun and Fan, Youheng and Huang, Xiaoyun and Tian, Lei},
	booktitle={2016 ASABE Annual International Meeting},
	pages={1},
	year={2016},
	organization={American Society of Agricultural and Biological Engineers}
}

@inproceedings{carrio2016ubristes,
	title={{UBRISTES: UAV-based building rehabilitation with visible and thermal infrared remote sensing}},
	author={Carrio, Adrian and Pestana, Jes{\'u}s and Sanchez-Lopez, Jose-Luis and Suarez-Fernandez, Ramon and Campoy, Pascual and Tendero, Ricardo and Garc{\'\i}a-De-Viedma, Mar{\'\i}a and Gonz{\'a}lez-Rodrigo, Beatriz and Bonatti, Javier and Rejas-Ayuga, Juan Gregorio and others},
	booktitle={Robot 2015: Second Iberian Robotics Conference},
	pages={245--256},
	year={2016},
	organization={Springer}
}

@article{olivares2015towards,
	title={Towards an autonomous vision-based unmanned aerial system against wildlife poachers},
	author={Olivares-Mendez, Miguel A and Fu, Changhong and Ludivig, Philippe and Bissyand{\'e}, Tegawend{\'e} F and Kannan, Somasundar and Zurad, Maciej and Annaiyan, Arun and Voos, Holger and Campoy, Pascual},
	journal={Sensors},
	volume={15},
	number={12},
	pages={31362--31391},
	year={2015},
	publisher={Multidisciplinary Digital Publishing Institute}
}

@inproceedings{martinez2014towards,
	title={Towards autonomous detection and tracking of electric towers for aerial power line inspection},
	author={Martinez, Carol and Sampedro, Carlos and Chauhan, Aneesh and Campoy, Pascual},
	booktitle={2014 International Conference on Unmanned Aircraft Systems (ICUAS)},
	pages={284--295},
	year={2014},
	organization={IEEE}
}

@article{hao2017wireless,
	title={Wireless fractal ultra-dense cellular networks},
	author={Hao, Yixue and Chen, Min and Hu, Long and Song, Jeungeun and Volk, Mojca and Humar, Iztok},
	journal={Sensors},
	volume={17},
	number={4},
	pages={841},
	year={2017},
	publisher={Multidisciplinary Digital Publishing Institute}
}

@article{ge2016wireless,
	title={Wireless fractal cellular networks},
	author={Ge, Xiaohu and Qiu, Yehong and Chen, Jiaqi and Huang, Meidong and Xu, Hui and Xu, Jing and Zhang, Wuxiong and Yang, Yang and Wang, Cheng-Xiang and Thompson, John},
	journal={IEEE Wireless Communications},
	volume={23},
	number={5},
	pages={110--119},
	year={2016},
	publisher={IEEE}
}

@article{butterworth2000base,
	title={Base station placement for in-building mobile communication systems to yield high capacity and efficiency},
	author={Butterworth, Keith S and Sowerby, Kevin W and Williamson, Allan G},
	journal={IEEE Transactions on Communications},
	volume={48},
	number={4},
	pages={658--669},
	year={2000},
	publisher={IEEE}
}

@inproceedings{wright1998optimization,
	title={Optimization methods for base station placement in wireless applications},
	author={Wright, Margaret H},
	booktitle={VTC'98. 48th IEEE Vehicular Technology Conference. Pathway to Global Wireless Revolution (Cat. No. 98CH36151)},
	volume={1},
	pages={387--391},
	year={1998},
	organization={IEEE}
}

@article{article,
	author = {Mohamed Amine, Ouamri},
	year = {2017},
	month = {05},
	pages = {},
	title = {Base Station Placement Optimization Using Genetic Algortithm},
	journal = {International Journal of Computer Aided Engineering and Technology},
	doi = {10.1504/IJCAET.2020.10006440}
}

@inproceedings{inproceedingsaerial1,
	title={Efficient 3D aerial base station placement considering users mobility by reinforcement learning},
	author={Ghanavi, Rozhina and Kalantari, Elham and Sabbaghian, Maryam and Yanikomeroglu, Halim and Yongacoglu, Abbas},
	booktitle={2018 IEEE Wireless Communications and Networking Conference (WCNC)},
	pages={1--6},
	year={2018},
	organization={IEEE}
}

@article{lee2015base,
	title={{Base station placement algorithm for large-scale LTE heterogeneous networks}},
	author={Lee, Seungseob and Lee, SuKyoung and Kim, Kyungsoo and Kim, Yoon Hyuk},
	journal={PloS one},
	volume={10},
	number={10},
	pages={e0139190},
	year={2015},
	publisher={Public Library of Science}
}

@inproceedings{inproceedingsaerial2,
	title={{Efficient 3-D Placement of an Aerial Base Station in Next Generation Cellular Networks}},
	author={Bor-Yaliniz, R Irem and El-Keyi, Amr and Yanikomeroglu, Halim},
	booktitle={2016 IEEE International Conference on Communications (ICC)},
	pages={1--5},
	year={2016},
	organization={IEEE}
}

@inproceedings{torHos2010algorithm,
	title={An algorithm for automatic base station placement in cellular network deployment},
	author={T{\"o}r{\H{o}}s, Istv{\'a}n and Fazekas, P{\'e}ter},
	booktitle={Meeting of the European Network of Universities and Companies in Information and Communication Engineering},
	pages={21--30},
	year={2010},
	organization={Springer}
}

@inproceedings{han2001genetic,
	title={Genetic approach with a new representation for base station placement in mobile communications},
	author={Han, Jin Kyu and Park, Byoung Seong and Choi, Yong Seok and Park, Han Kyu},
	booktitle={IEEE 54th Vehicular Technology Conference. VTC Fall 2001. Proceedings (Cat. No. 01CH37211)},
	volume={4},
	pages={2703--2707},
	year={2001},
	organization={IEEE}
}

@inproceedings{inproceedings,
	author = {Raisanen, Larry and Whitaker, Roger},
	year = {2003},
	month = {01},
	pages = {714-720},
	title = {Multi-objective Optimization in the Area Coverage Problems for Cellular Communication Networks: Evaluation of an Elitist Evolutionary Strategy.},
	journal = {Proceedings of the ACM Symposium on Applied Computing},
	doi = {10.1145/952532.952672}
}

@online{matGeom,
	title = {{Matlab geometry toolbox for 2D/3D geometric computing: }}, howpublished ={\url{https://github.com/mattools/matGeom}}, note = {Accessed: 2020-01-20}
}

@online{winprop,
	title = {{Winprop}}, howpublished={\url{https://altairhyperworks.com/product/feko/winprop-propagation-modeling}}, note = {Accessed: 2020-01-20}
}

@online{wirelessinsite,
	title = {{Wireless InSite}}, howpublished={\url{https://www.remcom.com/wireless-insite-em-propagation-software}}, note = {Accessed: 2020-01-20}
}

@online{huawei,
	title = {{Huawei 5G Wireless Network Planning Solution White Paper: }}, howpublished={\url{https://www-file.huawei.com/-/media/corporate/pdf/white%20paper/2018/5g_wireless_network_planing_solution_en.pdf?la=en-ch}}, note = {Accessed: 2020-01-20}
}

@inproceedings{simic2017coverage,
	title={Coverage and robustness of mm-Wave urban cellular networks: Multi-frequency HetNets are the 5G future},
	author={Simic, Ljiljana and Panda, Soumendra and Riihijarvi, Janne and Mahonen, Petri},
	booktitle={2017 14th Annual IEEE International Conference on Sensing, Communication, and Networking (SECON)},
	pages={1--9},
	year={2017},
	organization={IEEE}
}

@inproceedings{chen2018encoder,
	title={Encoder-decoder with atrous separable convolution for semantic image segmentation},
	author={Chen, Liang-Chieh and Zhu, Yukun and Papandreou, George and Schroff, Florian and Adam, Hartwig},
	booktitle={Proceedings of the European Conference on Computer Vision (ECCV)},
	pages={801--818},
	year={2018}
}

@article{Vienna5GSLS,
	title = {Flexible multi-node simulation of cellular mobile communications: the {Vienna 5G System Level Simulator}},
	author = {Martin Klaus Müller and Fjolla Ademaj and Thomas Dittrich and Agnes Fastenbauer and Blanca Ramos Elbal and Armand Nabavi and Lukas Nagel and Stefan Schwarz and Markus Rupp},
	journal = {EURASIP Journal on Wireless Communications and Networking},
	year = {2018},
	month = sep,
	volume = {2018},
	number = {1},
	pages = {17},
	doi = {10.1186/s13638-018-1238-7}
}

@misc{potsdam,
	title = {{ISPRS Test Project on Urban Classification and 3D Building Reconstruction}}, howpublished ={\url{http://www2.isprs.org/commissions/comm3/wg4/detection-and-reconstruction.html}}, note = {Accessed: 2020-01-22}
}

@article{deb2002fast,
	title={{A fast and elitist multiobjective genetic algorithm: NSGA-II}},
	author={Deb, Kalyanmoy and Pratap, Amrit and Agarwal, Sameer and Meyarivan, TAMT},
	journal={IEEE Transactions on Evolutionary Computation},
	volume={6},
	number={2},
	pages={182--197},
	year={2002},
	publisher={IEEE}
}

@inproceedings{farmbeats,
	title={Farmbeats: An IoT platform for data-driven agriculture},
	author={Vasisht, Deepak and Kapetanovic, Zerina and Won, Jongho and Jin, Xinxin and Chandra, Ranveer and Sinha, Sudipta and Kapoor, Ashish and Sudarshan, Madhusudhan and Stratman, Sean},
	booktitle={{14th USENIX Symposium on Networked Systems Design and Implementation (NSDI 17)}},
	pages={515--529},
	year={2017}
}

@misc{sunroof,
	title = {{Google Sunroof: }}, howpublished ={\url{https://www.google.com/get/sunroof}}, note = {Accessed: 2020-01-22}
}

@misc{dronesurvey,
	title={Aerial drone for well-site and signal survey},
	author={Whipple, John and Jeirath, Nakul and Archer, Cameron and Sisk, David Allen and Gray, Stephen and Lee, Cody James and Gonzalez, Jesus and Wilmes, Theodore and others},
	year={2019},
	month=jan # "~29",
	publisher={Google Patents},
	note={US Patent 10,192,182}
}

@article{dronesurvey2,
	title={{Mobile 3D mapping for surveying earthwork projects using an Unmanned Aerial Vehicle (UAV) system}},
	author={Siebert, Sebastian and Teizer, Jochen},
	journal={Automation in construction},
	volume={41},
	pages={1--14},
	year={2014},
	publisher={Elsevier}
}

@article{dronesurvey3,
	title={Unmanned aerial vehicles for surveying marine fauna: assessing detection probability},
	author={Hodgson, Amanda and Peel, David and Kelly, Natalie},
	journal={Ecological Applications},
	volume={27},
	number={4},
	pages={1253--1267},
	year={2017},
	publisher={Wiley Online Library}
}

@article{pool,
	title={An effective method for detecting potential woodland vernal pools using high-resolution LiDAR data and aerial imagery},
	author={Wu, Qiusheng and Lane, Charles and Liu, Hongxing},
	journal={Remote Sensing},
	volume={6},
	number={11},
	pages={11444--11467},
	year={2014},
	publisher={Multidisciplinary Digital Publishing Institute}
}

@article{landcover,
	title={Land cover classification from fused DSM and UAV images using convolutional neural networks},
	author={Al-Najjar, Husam AH and Kalantar, Bahareh and Pradhan, Biswajeet and Saeidi, Vahideh and Halin, Alfian Abdul and Ueda, Naonori and Mansor, Shattri},
	journal={Remote Sensing},
	volume={11},
	number={12},
	pages={1461},
	year={2019},
	publisher={Multidisciplinary Digital Publishing Institute}
}

@article{haklay2008openstreetmap,
	title={Openstreetmap: User-generated street maps},
	author={Haklay, Mordechai and Weber, Patrick},
	journal={IEEE Pervasive Computing},
	volume={7},
	number={4},
	pages={12--18},
	year={2008},
	publisher={Ieee}
}

@article{molina2017lte,
	title={{LTE-V for sidelink 5G V2X vehicular communications: A new 5G technology for short-range vehicle-to-everything communications}},
	author={Molina-Masegosa, Rafael and Gozalvez, Javier},
	journal={IEEE Vehicular Technology Magazine},
	volume={12},
	number={4},
	pages={30--39},
	year={2017},
	publisher={IEEE}
}

@article{wang2018holo3dgis,
	title={{Holo3DGIS: Leveraging Microsoft HoloLens in 3D geographic information}},
	author={Wang, Wei and Wu, Xingxing and Chen, Guanchen and Chen, Zeqiang},
	journal={ISPRS International Journal of Geo-Information},
	volume={7},
	number={2},
	pages={60},
	year={2018},
	publisher={Multidisciplinary Digital Publishing Institute}
}

@inproceedings{gupta2012nsga,
	title={{A NSGA-II based approach for camera placement problem in large scale surveillance application}},
	author={Gupta, Ankit and Pati, Kumar Ashis and Subramanian, Venkatesh K},
	booktitle={2012 4th International Conference on Intelligent and Advanced Systems (ICIAS2012)},
	volume={1},
	pages={347--352},
	year={2012},
	organization={IEEE}
}

@techreport{3gpp.36.942,
	author = {3GPP},
	day = {09},
	institution = {{3rd Generation Partnership Project (3GPP)}},
	month = {07},
	note = {Version 15.0.0},
	number = {36.942},
	title = {{Evolved Universal Terrestrial Radio Access (E-UTRA); Radio Frequency (RF) system scenarios}},
	type = {Technical report (TR)},
	url = {https://portal.3gpp.org/desktopmodules/Specifications/SpecificationDetails.aspx?specificationId=2592},
	year = {2018}
}

@misc{rappaport2006method,
	title={Method and system for designing or deploying a communications network which allows simultaneous selection of multiple components},
	author={Rappaport, Theodore and Skidmore, Roger},
	year={2006},
	month=aug # "~22",
	publisher={Google Patents},
	note={US Patent 7,096,173}
}

@misc{rappaport2001method,
	title={Method and system for automated optimization of antenna positioning in 3-D},
	author={Rappaport, Theodore S and Skidmore, Roger R},
	year={2001},
	month=nov # "~13",
	publisher={Google Patents},
	note={US Patent 6,317,599}
}

@misc{rappaport2007system,
	title={System and method for automated placement or configuration of equipment for obtaining desired network performance objectives},
	author={Rappaport, Theodore S and Skidmore, Roger},
	year={2007},
	month=nov # "~13",
	publisher={Google Patents},
	note={US Patent 7,295,960}
}

@misc{ephremides1999method,
	title={Method and device for placement of transmitters in wireless networks},
	author={Ephremides, Anthony and Stamatelos, Dimitrios},
	year={1999},
	month=nov # "~16",
	publisher={Google Patents},
	note={US Patent 5,987,328}
}

@misc{cutrer1997measurement,
	title={{Measurement-based method of optimizing the placement of antennas in a RF distribution system}},
	author={Cutrer, David M and Georges, John B and Lau, Kam Y},
	year={1997},
	month=sep # "~16",
	publisher={Google Patents},
	note={US Patent 5,668,562}
}

@misc{inasawamethod,
	title={Method for deciding optimum layout of indoor base station and optimum layout decision system},
	year={2000},
	author={Inasawa, Yoshio and Chiba, Isamu},
	note={Japanese Patent 2000333239}
}

@article{sun2017nyusim,
	title={NYUSIM User Manual},
	author={Sun, Shu},
	journal={New York University and NYU WIRELESS},
	year={2017}
}

@inproceedings{kamnitsas2017ensembles,
	title={Ensembles of multiple models and architectures for robust brain tumour segmentation},
	author={Kamnitsas, Konstantinos and Bai, Wenjia and Ferrante, Enzo and McDonagh, Steven and Sinclair, Matthew and Pawlowski, Nick and Rajchl, Martin and Lee, Matthew and Kainz, Bernhard and Rueckert, Daniel and others},
	booktitle={International MICCAI Brainlesion Workshop},
	pages={450--462},
	year={2017},
	organization={Springer}
}

@article{dumoulin2016guide,
	title={A guide to convolution arithmetic for deep learning},
	author={Dumoulin, Vincent and Visin, Francesco},
	journal={arXiv preprint arXiv:1603.07285},
	year={2016}
}

@inproceedings{long2015fully,
	title={Fully convolutional networks for semantic segmentation},
	author={Long, Jonathan and Shelhamer, Evan and Darrell, Trevor},
	booktitle={Proceedings of the IEEE Conference on Computer Vision and Pattern Recognition},
	pages={3431--3440},
	year={2015}
}

@inproceedings{ronneberger2015u,
	title={U-net: Convolutional Networks for Biomedical Image Segmentation},
	author={Ronneberger, Olaf and Fischer, Philipp and Brox, Thomas},
	booktitle={International Conference on Medical Image Computing and Computer-assisted Intervention},
	pages={234--241},
	year={2015},
	organization={Springer}
}

@inproceedings{zhao2017pyramid,
	title={Pyramid scene parsing network},
	author={Zhao, Hengshuang and Shi, Jianping and Qi, Xiaojuan and Wang, Xiaogang and Jia, Jiaya},
	booktitle={Proceedings of the IEEE conference on Computer Vision and Pattern Recognition},
	pages={2881--2890},
	year={2017}
}

@article{zhou2014object,
	title={{Object Detectors Emerge in Deep Scene CNNs}},
	author={Zhou, Bolei and Khosla, Aditya and Lapedriza, Agata and Oliva, Aude and Torralba, Antonio},
	journal={arXiv preprint arXiv:1412.6856},
	year={2014}
}

@misc{nvidiaDriving,
	title = {{Pixel-Perfect Perception: How AI Helps Autonomous Vehicles See Outside the Box}}, howpublished ={\url{https://blogs.nvidia.com/blog/2019/10/23/drive-labs-panoptic-segmentation/}}, note = {Accessed: 2020-01-22}
}

@inproceedings{yeboah2018semantic,
	title={Semantic scene segmentation for indoor robot navigation via deep learning},
	author={Yeboah, Yao and Yanguang, Cai and Wu, Wei and Farisi, Zeyad},
	booktitle={Proceedings of the 3rd International Conference on Robotics, Control and Automation},
	pages={112--118},
	year={2018}
}

@inproceedings{chollet2017xception,
	title={Xception: Deep learning with depthwise separable convolutions},
	author={Chollet, Fran{\c{c}}ois},
	booktitle={Proceedings of the IEEE Conference on Computer Vision and Pattern Recognition},
	pages={1251--1258},
	year={2017}
}

@inproceedings{he2016deep,
	title={Deep residual learning for image recognition},
	author={He, Kaiming and Zhang, Xiangyu and Ren, Shaoqing and Sun, Jian},
	booktitle={Proceedings of the IEEE Conference on Computer Vision and Pattern Recognition},
	pages={770--778},
	year={2016}
}

@inproceedings{deng2009imagenet,
	title={Imagenet: A large-scale hierarchical image database},
	author={Deng, Jia and Dong, Wei and Socher, Richard and Li, Li-Jia and Li, Kai and Fei-Fei, Li},
	booktitle={2009 IEEE conference on Computer Vision and Pattern Recognition},
	pages={248--255},
	year={2009},
	organization={Ieee}
}

@inproceedings{csurka2013good,
	title={What is a good evaluation measure for semantic segmentation?.},
	author={Csurka, Gabriela and Larlus, Diane and Perronnin, Florent and Meylan, France},
	booktitle={BMVC},
	volume={27},
	pages={2013},
	year={2013}
}
